\documentclass[english,preprint]{aastex}
\usepackage[T1]{fontenc}
\usepackage[latin9]{inputenc}
\usepackage{amsmath}
\usepackage{amssymb}

\makeatletter
\usepackage{amsmath}



\makeatother

\usepackage{babel}
\begin{document}

\title{Characterizing Magnetized Turbulence in M51}

\author{Martin Houde$^{1,2}$, Andrew Fletcher$^{3}$, Rainer Beck$^{4}$,
Roger H. Hildebrand$^{5,6}$, John E. Vaillancourt$^{7}$, and Jeroen
M. Stil$^{8}$ }

\affil{$^{1}$Department of Physics and Astronomy, The University of Western
Ontario, London, ON, N6A 3K7, Canada}

\affil{$^{2}$Division of Physics, Mathematics and Astronomy, California
Institute of Technology, Pasadena, CA 91125 }

\affil{$^{3}$School of Mathematics and Statistics, Newcastle University,
Newcastle-upon-Tyne NE1 7RU, U.K.}

\affil{$^{4}$Max-Planck-Institut f\"{u}r Radioastronomie, Auf dem H\"{u}gel
69, 53121 Bonn, Germany}

\affil{$^{5}$Department of Astronomy and Astrophysics and Enrico Fermi
Institute, The University of Chicago, Chicago, IL 60637}

\affil{$^{6}$Department of Physics, The University of Chicago, Chicago,
IL 60637}

\affil{$^{7}$Stratospheric Observatory for Infrared Astronomy, Universities
Space Research Association, NASA Ames Research Center, Moffet Field,
CA 94035}

\affil{$^{8}$Department of Physics and Astronomy, The University of Calgary,
Calgary, AB, T2N 1N4, Canada}
\begin{abstract}
We use previously published high-resolution synchrotron polarization
data to perform an angular dispersion analysis with the aim of charactering
magnetized turbulence in M51. We first analyze three distinct regions
(the center of the galaxy, and the northwest and southwest spiral
arms) and can clearly discern the turbulent correlation length scale
from the width of the magnetized turbulent correlation function for
two regions and detect the imprint of anisotropy in the turbulence
for all three. Furthermore, analyzing the galaxy as a whole allows
us to determine a two-dimensional Gaussian model for the magnetized
turbulence in M51. We measure the turbulent correlation scales parallel
and perpendicular to the local mean magnetic field to be, respectively,
$\delta_{\Vert}=98\pm5$ pc and $\delta_{\bot}=54\pm3$ pc, while
the turbulent to ordered magnetic field strength ratio is found to
be $B_{\mathrm{t}}/B_{0}=1.01\pm0.04$. These results are consistent
with those of \citet{Fletcher2011}, who performed a Faraday rotation
dispersion analysis of the same data, and our detection of anisotropy
is consistent with current magnetized turbulence theories. 
\end{abstract}


\section{Introduction}

The magnetized diffuse interstellar medium, of the Milky Way and other
galaxies, is turbulent and so an understanding of its properties and
role should include the quantities commonly used to describe turbulence:
characteristic length scales, power spectra, the relative energies
in the mean and fluctuating components, and so on. One important property
of magnetohydrodynamic (MHD) turbulence is that the random fluctuations
in the inertial range are not necessarily isotropic, as is the case
in the classical picture of purely hydrodynamic, incompressible Kolmogorov
turbulence: the correlation length of magnetic fluctuations can be
larger along the mean field direction compared to the perpendicular
direction. This mean field can be either an external large-scale magnetic
field or simply the magnetic field at the largest scale of a turbulent
eddy acting on fluctuations within the eddy on smaller scales. As
well as the inherent anisotropy of MHD turbulence dynamical effects
in the ISM flow such as shocks and shear, due to localized sources
such as supernovae or global features like differential rotation,
can also imprint anisotropy on the turbulence. 

There have been a few indications from observations that magnetic
field fluctuations in the ISM exhibit anisotropies. \citet{Brown2001}
binned Faraday rotation measures (RM) for extra-galactic (EG) sources
in the Galactic plane in the range $82\deg<l<146\deg$ and found that
the variance in a bin is correlated with the magnitude of the mean
RM; higher RMs are associated with stronger fluctuations, and it was
proposed that this occurs because the fluctuations in the magnetic
field are mainly aligned with the mean field. \citet{Jaffe2010} fitted
a model magnetic field to the synchrotron emission and EG RMs along
the Galactic plane and found that an anisotropic (or in their terminology
an ordered random) magnetic field component was required to fit the
observations along with both a mean field and an isotopic random magnetic
field. Similarly, \citet{Jansson2012} required either an anisotropic
(or striated field in their terminology) magnetic field component,
or a correlation between the mean magnetic field and cosmic ray density,
to obtain good fits to all-sky observations of synchrotron emission
and RMs (in their model anisotropic random fields and close cosmic
ray-to-mean magnetic field coupling are degenerate parameters). Away
from the Milky Way, \citet{Beck2005} compared the observed increase
in both total and polarized synchrotron emission at the strong shock
fronts along the bars of the galaxies NGC 1097 and NGC 1365 with theoretical
expectations based on the compression and shear of random and mean
magnetic fields; their results indicate that strong anisotropic random
magnetic fields are produced at these positions. \citet{Fletcher2011}
attributed the order of magnitude difference between ordered magnetic
field strengths obtained via equipartition estimates and Faraday rotation
modeling, to a strong anisotropic random magnetic field in the nearby
galaxy M51: this component is responsible for the strong polarized
signal but contributes little to the Faraday rotation.

\begin{figure}
\epsscale{0.75}\plotone{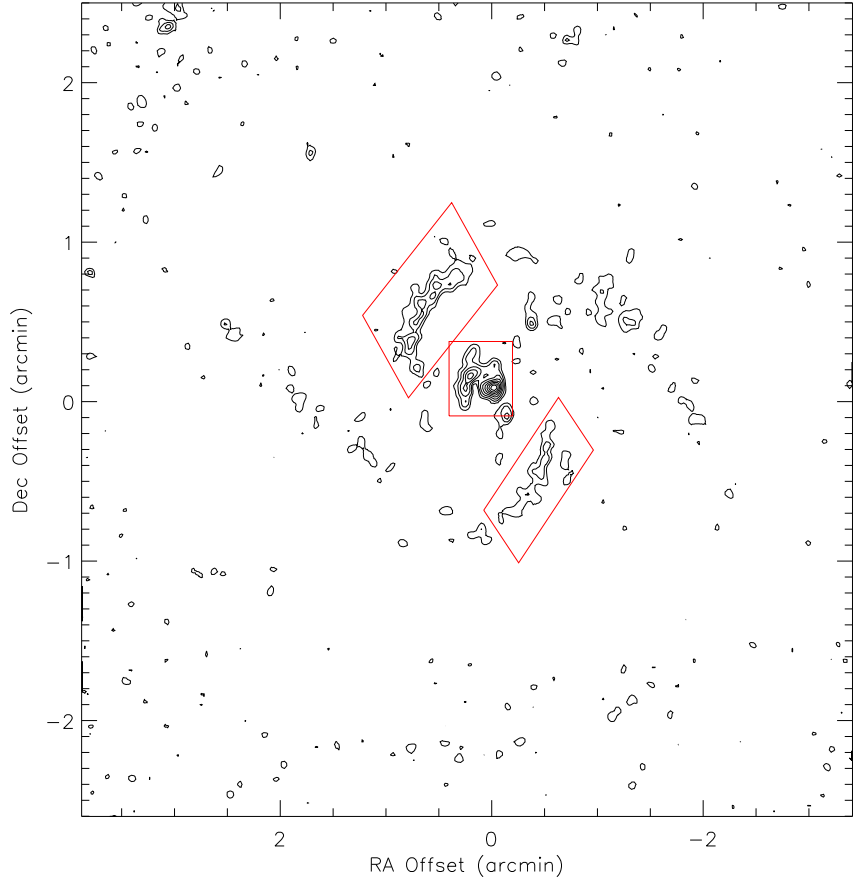}

\caption{\label{fig:m51_polflux}Polarized flux at $\lambda6.2$ cm for M51.
There are three regions that can be independently used (or combined)
for a dispersion analysis: the spiral arms in the northeast and southwest,
and the center of the galaxy. These regions are contained within the
corresponding three red rectangles in the figure. The map is centered
at\emph{ }$\mathrm{RA\,}(\mathrm{J2000})=11^{\mathrm{h}}29^{\mathrm{m}}52\fs4$,
$\mathrm{Dec\,}(\mathrm{J2000})=47^{\circ}11\arcmin43\farcs5$, and
the contours are drawn at 20 to 80 percent (10 percent increments)
of the peak polarized flux density ($173\,\mu$Jy/beam).}
\end{figure}

In this paper we perform an angular dispersion analysis, based on
the work of \citet{Falceta2008}, \citet{Hildebrand2009}, and \citet{Houde2009,Houde2011},
on the Effelsberg 100-m/VLA $\lambda6.2$ cm synchrotron polarization
map of \citet{Fletcher2011} ($4\arcsec$ FWHM resolution and $1\arcsec$
sampling). We show in Figure \ref{fig:m51_polflux} the global view
of M51 in polarized flux provided by these data. There are clearly
only three regions that can be used, or combined, for a dispersion
analysis: the spiral arms in the northeast and southwest, and the
center of the galaxy. These regions are contained within the corresponding
three red rectangles in the figure. 

Although these data were obtained with high spatial resolution, they
will not allow us to get a handle on the magnetized turbulence power
spectrum as done in \citet{Houde2011} for Galactic molecular clouds
(note that $1\arcsec\simeq37$ pc). Nonetheless, \citet{Fletcher2011}
calculate from a Faraday rotation dispersion analysis that the size
of a turbulence cell should be approximately 50 pc. This suggests
that we may be able to measure and determine this value independently
with the data at hand, given the expected size of a cell and the aforementioned
spatial resolution. We would be in a position to not only determine
the number of turbulence cells contained in the average column of
gas subtended by the telescope beam, but also get accurate values
for the ratio of the turbulent-to-ordered magnetic energy in different
parts of M51. Furthermore, in view of the large numbers of polarization
measurements available with this map the statistics may be good enough
to allow a study of possible anisotropy in the autocorrelation function
of magnetized turbulence.

We start in the next section by first giving a summary of the dispersion
analysis in Section \ref{sec:Analysis}, a description of the data
used for our analysis is given in Section \ref{sec:Observations},
which is then followed in Section \ref{sub:Isotropic-Turbulence}
by an isotropic dispersion analysis on the three available regions
in the manner presented in \citet{Houde2009,Houde2011} and \citet{Hildebrand2009}.
A first attempt at measuring any potential anisotropy is presented
in Section \ref{sub:hybrid} through the independent analyses on displacement
vectors that are grouped into two sets, which are either oriented
approximately parallel or perpendicular to the local mean magnetic
field. The derived turbulence autocorrelation functions can then be
compared and any differences in their widths will reveal an anisotropy.
Finally, in Section \ref{sub:Two-dimensional Anisotropic} we apply
the dispersion analysis to M51 as a whole (i.e., by simultaneously
using all available polarization vectors) to map the two-dimensional
turbulence autocorrelation function to once again reveal any anisotropy
in the magnetized turbulence, but in more detail. We finish with a
discussion of our results in Section \ref{sec:Discussion}, a brief
summary in Section \ref{sub:Summary}, while more details concerning
the dispersion analysis will be found in the Appendix.

\section{Angular Dispersion Analysis \label{sec:Analysis}}

Structure functions have long been used in physics and astrophysics
to characterize turbulence, as they allow for the treatment of power-law
energy spectra, such as those found in Kolmogorov turbulence, without
the mathematical divergences associated with stationary signal \citep{Frisch1995, Beck1999, Falceta2008}.
Such structure functions, of varying orders, can be calculated for
a range of physical parameters (e.g., velocity and density fields).
In this paper, we intend to apply the angular dispersion analysis
previously introduced in the literature \citep{Falceta2008,Hildebrand2009,Houde2009,Houde2011},
where the chosen parameter is the orientation of the projection of
the magnetic field on the plane of the sky. More precisely, we will
use the polarization angle orientation in lieu of that of the magnetic
field. For the polarization of synchrotron (or dust) emission this
angle is orientated at $90^{\circ}$ from that of the projected magnetic
field. Although we will provide a summary of the important relations
required for the angular dispersion analysis later in Section \ref{sub:Angular-Dispersion-Function},
a simplified exposition based on material that can be found in \citet{Hildebrand2009}
is first given here.

\subsection{Angular Structure Function\label{sub:Angular-Structure-Function}}

We start by defining the difference $\Delta\Phi\left(\boldsymbol{\ell}\right)$
in the orientation of the magnetic field (unless otherwise specified,
in this paper we will only concern ourselves with the plane of the
sky component of the magnetic field) at two points separated by a
distance $\boldsymbol{\ell}$ 

\begin{equation}
\Delta\Phi\left(\boldsymbol{\ell}\right)\equiv\Phi\left(\mathbf{r}\right)-\Phi\left(\mathbf{r}+\boldsymbol{\ell}\right),\label{eq:Dphi}
\end{equation}

\noindent with $\Phi\left(\mathbf{r}\right)$ the magnetic field orientation
at position $\mathbf{r}$ (both $\mathbf{r}$ and $\boldsymbol{\ell}$
are also understood to be located in the plane of the sky). Given
a set of measurements $\Phi\left(\mathbf{r}\right)$ on a polarization
map, we can also define the following (second order) angular structure
function

\begin{equation}
\left\langle \Delta\Phi^{2}\left(\ell\right)\right\rangle \equiv\frac{1}{N\left(\ell\right)}\sum_{i=1}^{N\left(\ell\right)}\left[\Phi\left(\mathbf{r}\right)-\Phi\left(\mathbf{r}+\boldsymbol{\ell}\right)\right]^{2}\label{eq:SF}
\end{equation}

\noindent where $\left\langle \cdots\right\rangle $ denotes an average,
$N\left(\ell\right)$ is the number of pairs of field orientation
measurements separated by $\ell=\left|\boldsymbol{\ell}\right|$,
and stationarity and isotropy were assumed (i.e., the structure function
is only dependent on the magnitude of $\boldsymbol{\ell}$, and not
on its orientation or $\mathbf{r}$; see \citealt{Falceta2008,Hildebrand2009}). 

\begin{figure}
\epsscale{0.8}\plotone{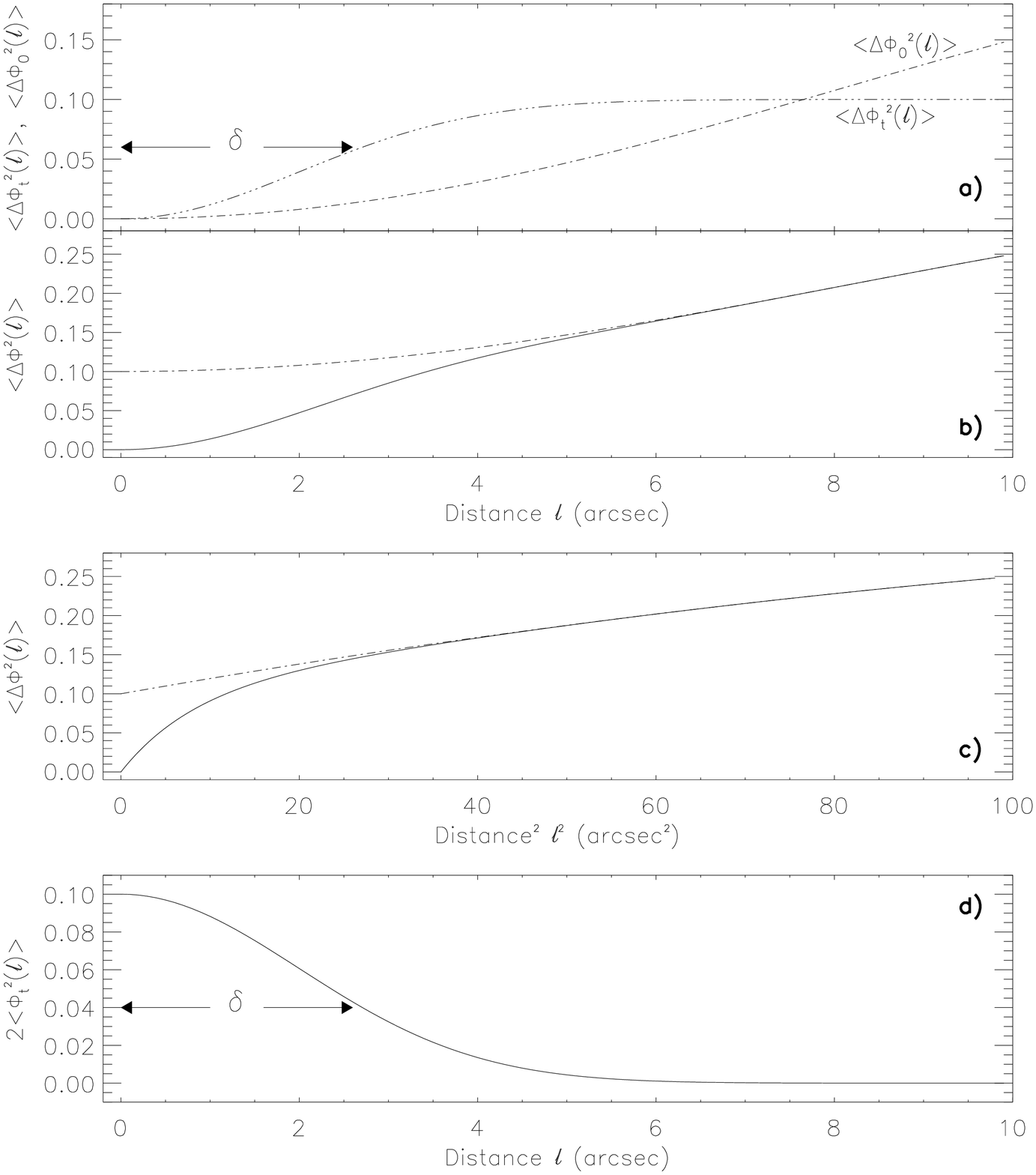}

\caption{\label{fig:SF}An example of an angular structure function. Panel
a) shows hypothetical turbulent $\left\langle \Delta\Phi_{\mathrm{t}}^{2}\left(\ell\right)\right\rangle $
(triple-dot-broken curve) and ordered $\left\langle \Delta\Phi_{0}^{2}\left(\ell\right)\right\rangle $
(dot-broken curve) contributions to the total angular structure function
$\left\langle \Delta\Phi^{2}\left(\ell\right)\right\rangle $ plotted
in Panel b) (solid curve). Panel c) presents the same information
as Panel b) but displayed as a function of $\ell^{2}$ instead of
$\ell$. The total angular structure function is subtracted from a
Taylor series fit obtained from its values at $\ell\geq6\arcsec$
(dot-broken curve in Panels b) and c)) to get the turbulent autocorrelation
function shown in Panel d). The correlation length characterizing
turbulence is represented by $\delta$.}
\end{figure}

The main assumption in our analysis consists in modeling the magnetic
field, and therefore its orientation through $\Phi\left(\mathbf{r}\right)$,
as being composed of a large-scale, ordered component $\Phi_{0}\left(\mathbf{r}\right)$
and a smaller scale, zero-mean, turbulent component $\Phi_{\mathrm{t}}\left(\mathbf{r}\right)$.
That is, we write

\begin{equation}
\Phi\left(\mathbf{r}\right)=\Phi_{\mathrm{t}}\left(\mathbf{r}\right)+\Phi_{0}\left(\mathbf{r}\right),\label{eq:Phi_tot}
\end{equation}

\noindent which, if we further assume these two components to be statistically
independent, leads to

\begin{equation}
\left\langle \Delta\Phi^{2}\left(\ell\right)\right\rangle =\left\langle \Delta\Phi_{\mathrm{t}}^{2}\left(\ell\right)\right\rangle +\left\langle \Delta\Phi_{0}^{2}\left(\ell\right)\right\rangle .\label{eq:DPhi_tot}
\end{equation}

\noindent We thus find that the structure function is composed of
two angular components stemming from the contributions of the turbulent
and ordered magnetic fields. Our assumption on the difference between
the two scales therefore allows for their separation and analyses.
This is exemplified in Figure \ref{fig:SF} where we show (Panel a))
hypothetical turbulent and ordered contributions to the total angular
structure function (Panel b), solid curve), as expressed in Equation
(\ref{eq:DPhi_tot}). Panel c) presents the same information as Panel
b) for $\left\langle \Delta\Phi^{2}\left(\ell\right)\right\rangle $
but displayed as a function of $\ell^{2}$ instead of $\ell$. This
is to show that the separation of the two length scales is sometimes
easier to visualize, through the abrupt change in the slope of $\left\langle \Delta\Phi^{2}\left(\ell\right)\right\rangle $,
by using the square of the distance for the abscissa; we will use
both representations for the data analyzed later in this paper. The
total angular structure function (of Panels b) or c)) is the input
to our problem as obtained from a polarization map, which we seek
to analyze in order to characterize magnetized turbulence.

The behavior of turbulent and ordered contributions to the total angular
structure function of Panel a) can be qualitatively understood from
the fact that, evidently, they must equal zero when $\ell=0$ and
then initially increase with $\ell$. The turbulent structure function
will keep increasing until it reaches values for $\ell$ that sufficiently
exceed the turbulence correlation length $\delta$, at which point
it will reach its maximum value. This can also be understood more
quantitatively with the following relation

\begin{eqnarray}
\left\langle \Delta\Phi_{\mathrm{t}}^{2}\left(\ell\right)\right\rangle  & = & \left\langle \left[\Phi_{\mathrm{t}}\left(\mathbf{r}\right)-\Phi_{\mathrm{t}}\left(\mathbf{r}+\boldsymbol{\ell}\right)\right]^{2}\right\rangle \nonumber \\
 & = & 2\left[\left\langle \Phi_{\mathrm{t}}^{2}\left(0\right)\right\rangle -\left\langle \Phi_{\mathrm{t}}^{2}\left(\ell\right)\right\rangle \right],\label{eq:Dphi_turb}
\end{eqnarray}

\noindent where the (stationary and isotropic) turbulent autocorrelation
function is defined with

\begin{equation}
\left\langle \Phi_{\mathrm{t}}^{2}\left(\ell\right)\right\rangle \equiv\left\langle \Phi_{\mathrm{t}}\left(\mathbf{r}\right)\Phi_{\mathrm{t}}\left(\mathbf{r}+\boldsymbol{\ell}\right)\right\rangle .\label{eq:Phi_autoco}
\end{equation}

\noindent It therefore follows that $\left\langle \Delta\Phi_{\mathrm{t}}^{2}\left(\ell\right)\right\rangle =2\left\langle \Phi_{\mathrm{t}}^{2}\left(0\right)\right\rangle $
when $\ell\gg\delta$, as the turbulence is not correlated on such
scales and its autocorrelation vanishes. The ordered structure function
is expected to rise, at first monotonically, with increasing values
of $\ell$ in view of its larger-scale nature. For the example of
Figure \ref{fig:SF} we have characterized the turbulent component
with a Gaussian autocorrelation function of width, or correlation
length (i.e., its standard deviation) $\delta=2\arcsec$, while the
ordered structure function was modeled with a Taylor expansion in
powers of $\ell^{2}$. This restriction to even powers in $\ell$
is dictated from the assumption of isotropy for the structure function. 

With the previous assumption in the difference of the two length scales
it becomes possible to model the ordered component $\left\langle \Delta\Phi_{0}^{2}\left(\ell\right)\right\rangle $
independently of $\left\langle \Delta\Phi_{\mathrm{t}}^{2}\left(\ell\right)\right\rangle $
by using values of $\ell$ sufficiently large (i.e., sufficiently
greater than $\delta$) where any variation in $\left\langle \Delta\Phi_{\mathrm{t}}^{2}\left(\ell\right)\right\rangle $
is negligible. For our example, we chose $\ell\geq6\arcsec$ to obtain
the Taylor series fit given by the dot-broken curve in Panels b) and
c) of Figure \ref{fig:SF}. This curve is then representative of $\left\langle \Delta\Phi_{0}^{2}\left(\ell\right)\right\rangle $
but shifted up by the constant level of the turbulent component present
in that range. More precisely if we define a function $\chi\left(\ell\right)$
for this curve, then we write

\begin{equation}
\chi\left(\ell\right)=2\left\langle \Phi_{\mathrm{t}}^{2}\left(0\right)\right\rangle +\left\langle \Delta\Phi_{0}^{2}\left(\ell\right)\right\rangle .\label{eq:chi(ell)}
\end{equation}

\noindent In this paper we will focus on characterizing magnetized
turbulence and we are thus interested in isolating the turbulent component
of the structure function, or, alternatively, its autocorrelation.
The latter (multiplied by a factor of two) is readily evaluated from
Equations (\ref{eq:DPhi_tot}), (\ref{eq:Dphi_turb}), and (\ref{eq:chi(ell)})
with 

\begin{equation}
2\left\langle \Phi_{\mathrm{t}}^{2}\left(\ell\right)\right\rangle =\chi\left(\ell\right)-\left\langle \Delta\Phi^{2}\left(\ell\right)\right\rangle \label{eq:2Phi_autoco}
\end{equation}

\noindent and shown in Panel d) of Figure \ref{fig:SF} (i.e., as
the subtraction of the solid curve from the dot-broken curve in Panel
b)). 

Although we could very well use the angular structure function in
the manner presented in this section for our analyses of the M51 data,
we will nonetheless for the rest of our discussion focus instead on
the \emph{angular dispersion function }$1-\left\langle \cos\left[\Delta\Phi\left(\ell\right)\right]\right\rangle $.
We note, however, that the properties, method, and technique discussed
thus far for the structure function apply just as well to the dispersion
function. In fact, we note that the two angular functions are simply
related through

\begin{equation}
1-\left\langle \cos\left[\Delta\Phi\left(\ell\right)\right]\right\rangle \simeq\frac{1}{2}\left\langle \Delta\Phi^{2}\left(\ell\right)\right\rangle \label{eq:cos_app}
\end{equation}

\noindent when $\Delta\Phi\left(\ell\right)\ll1$. The advantage of
the dispersion function is its close connection to the autocorrelation
of the magnetic field (see Equation (\ref{eq:cos}) below), which
then naturally leads to the study of the magnetized turbulent power
spectrum through a simple Fourier transform \citep{Houde2011}.

Finally, it is important to note that in general the width of the
turbulent autocorrelation function (and of the associated structure/dispersion
function) is not solely due to the intrinsic correlation length of
turbulence. A correlation scale brought about by the finite spatial
resolution with which observations are realized will also combine
to the intrinsic turbulent correlation length to set the overall width
of the turbulent autocorrelation function. A further complication
results from the related problem of signal integration through the
line of sight. As we will soon see, a careful analysis of such effects
will not only allow us to disentangle the intrinsic correlation length
characterizing magnetized turbulence from the overall width of the
turbulent autocorrelation function, but also to determine the level
of turbulent energy contained in the medium probed by the observations.
For this to be feasible, however, some approximation must be made
on the nature of the turbulent autocorrelation function. The case
of isotropic Gaussian turbulence and beam profile functions was treated
in the details in \citet{Houde2009} (see their Section 2) and their
main relations detailing the combination of the two length scales
for the analysis of turbulence are given in the next section (see
Equations (\ref{eq:b^2(ell)})-(\ref{Taylor_iso}) below). In this
paper, we further provide an analysis of the more general case of
anisotropic Gaussian turbulence in the Appendix, while the corresponding
results are also summarized in Section \ref{sub:Angular-Dispersion-Function}
(see Equations (\ref{eq:b^2_inclined})-(\ref{mu2})). This will in
turn make possible the measurement of anisotropy in the turbulent
autocorrelation function, which is another important parameter for
the characterization of magnetized turbulence.

\subsection{Angular Dispersion Function\label{sub:Angular-Dispersion-Function} }

As previously mentioned, the analysis of the angular dispersion function
found in the Appendix and summarized in this section follows that
presented in Section 2 of \citet{Houde2009} with the difference that
we now allow for the presence of anisotropy in the turbulence. As
stated in Section \ref{sub:Angular-Structure-Function}, we are interested
in the function $\cos\left[\Delta\Phi\left(\boldsymbol{\ell}\right)\right]$
that is related to the magnetic field autocorrelation function through

\begin{equation}
\left\langle \cos\left[\Delta\Phi\left(\boldsymbol{\ell}\right)\right]\right\rangle =\frac{\left\langle \overline{\mathbf{B}}\mathbf{\cdot}\overline{\mathbf{B}}\mathbf{\left(\boldsymbol{\ell}\right)}\right\rangle }{\left\langle \overline{\mathbf{B}}\mathbf{\cdot}\overline{\mathbf{B}}\left(0\right)\right\rangle },\label{eq:cos}
\end{equation}

\noindent where $\left\langle \cdots\right\rangle $ denotes an average
and $\left\langle \overline{\mathbf{B}}\mathbf{\cdot}\overline{\mathbf{B}}\mathbf{\left(\boldsymbol{\ell}\right)}\right\rangle \equiv\left\langle \overline{\mathbf{B}}\left(\mathbf{r}\right)\mathbf{\cdot}\overline{\mathbf{B}}\mathbf{\left(\mathbf{r}+\boldsymbol{\ell}\right)}\right\rangle $.
It is important to note that the magnetic field $\overline{\mathbf{B}}$
is a weighted average (with the polarized flux) through the thickness
of the column of gas probed (i.e., the disk of M51) and across the
telescope beam (see Equation (\ref{eq:Bbar})). The local, non-averaged,
magnetic field $\mathbf{B\left(\mathbf{x}\right)}$ at a point $\mathbf{x}$
is composed of an ordered field $\mathbf{B}_{0}\mathbf{\left(\mathbf{x}\right)}$
and a turbulent component $\mathbf{B_{\mathrm{t}}\left(\mathbf{x}\right)}$
such that 

\begin{equation}
\mathbf{\mathbf{B\left(\mathbf{x}\right)}}=\mathbf{B}_{0}\mathbf{\mathbf{\mathbf{\left(\mathbf{x}\right)+\mathbf{B}_{\mathrm{t}}\left(\mathbf{x}\right)}}}.\label{eq:Btot}
\end{equation}

As was the case earlier, the displacement vector $\boldsymbol{\ell}$
in Equation (\ref{eq:cos}), and others that will follow, is understood
to be located in the plane of the sky. We will further break $\boldsymbol{\ell}$
down into two perpendicular components 

\begin{equation}
\boldsymbol{\ell}=\boldsymbol{\ell}_{1}+\boldsymbol{\ell}_{2},\label{eq:ell}
\end{equation}

\noindent where, unless otherwise noted, $\boldsymbol{\ell}_{1}$
and $\boldsymbol{\ell}_{2}$ are taken to be respectively perpendicular
and parallel to the projection of the ordered component of the magnetic
field $\mathbf{B}_{0}$ on the plane of the sky. For everything that
follows statistical independence between $\mathbf{B}_{0}\mathbf{\left(\mathbf{x}\right)}$
and $\mathbf{B_{\mathrm{t}}\left(\mathbf{x}\right)}$, homogeneity
in the strength of the magnetic fields $\left\langle B_{0}^{2}\right\rangle \equiv\left\langle \mathbf{B}_{0}\mathbf{\cdot}\mathbf{B}_{0}\left(0\right)\right\rangle $
and $\left\langle B_{\mathrm{t}}^{2}\right\rangle \equiv\left\langle \mathbf{B}_{\mathrm{t}}\mathbf{\cdot}\mathbf{B}_{\mathrm{t}}\left(0\right)\right\rangle $,
as well as, more generally, stationarity in the autocorrelation functions
$\left\langle \mathbf{B}_{0}\mathbf{\cdot}\mathbf{B}_{0}\mathbf{\left(\boldsymbol{\ell}\right)}\right\rangle $
and $\left\langle \mathbf{B}_{\mathrm{t}}\mathbf{\cdot}\mathbf{B}_{\mathrm{t}}\mathbf{\left(\boldsymbol{\ell}\right)}\right\rangle $
are assumed. The assumption of homogeneity in the field strength in
particular, while clearly an idealization, is needed for securing
a quantitative measure of turbulence from our data (see the Appendix
for more details). 

Using these assumptions it can be shown that, just as was the case
for the structure function in Equation (\ref{eq:DPhi_tot}), the dispersion
function $1-\left\langle \cos\left[\Delta\Phi\left(\boldsymbol{\ell}\right)\right]\right\rangle $
can be decomposed into turbulent and ordered terms 

\begin{eqnarray}
1-\left\langle \cos\left[\Delta\Phi\left(\boldsymbol{\ell}\right)\right]\right\rangle  & = & \left[b^{2}\left(0\right)-b^{2}\left(\boldsymbol{\ell}\right)\right]+\left[\alpha^{2}\left(0\right)-\alpha^{2}\left(\boldsymbol{\ell}\right)\right]\nonumber \\
 & = & \left\{ b^{2}\left(0\right)+\left[\alpha^{2}\left(0\right)-\alpha^{2}\left(\boldsymbol{\ell}\right)\right]\right\} -b^{2}\left(\boldsymbol{\ell}\right),\label{eq:b_alpha}
\end{eqnarray}

\noindent with the normalized ordered and turbulent autocorrelation
functions given by 

\begin{eqnarray}
\alpha^{2}\left(\boldsymbol{\ell}\right) & = & \frac{\left\langle \overline{\mathbf{B}}_{0}\mathbf{\cdot}\overline{\mathbf{B}}_{0}\left(\boldsymbol{\ell}\right)\right\rangle }{\left\langle \overline{\mathbf{B}}\mathbf{\cdot}\overline{\mathbf{B}}\left(0\right)\right\rangle }\label{eq:alpha}\\
b^{2}\left(\boldsymbol{\ell}\right) & = & \frac{\left\langle \overline{\mathbf{B}}_{\mathrm{t}}\mathbf{\cdot}\overline{\mathbf{B}}_{\mathrm{t}}\left(\boldsymbol{\ell}\right)\right\rangle }{\left\langle \overline{\mathbf{B}}\mathbf{\cdot}\overline{\mathbf{B}}\left(0\right)\right\rangle },\label{eq:b2}
\end{eqnarray}

\noindent respectively. The quantity $b^{2}\left(0\right)$ in Equation
(\ref{eq:b_alpha}) is simply, from Equation (\ref{eq:b2}), the integrated
turbulent to total magnetic energy ratio. It is also the equivalent
of $\left\langle \Delta\Phi_{\mathrm{t}}^{2}\left(\ell\right)\right\rangle $
(see Equation (\ref{eq:Phi_autoco})) when dealing with the angular
structure function. The ordered function $\alpha^{2}\left(0\right)-\alpha^{2}\left(\boldsymbol{\ell}\right)$,
which we assume to be of a larger spatial scale than $b^{2}\left(\boldsymbol{\ell}\right)$,
can be advantageously modeled with a Taylor series. Since, as we will
soon discuss, we adopt a model of turbulence where the autocorrelation
function is even in directions parallel and perpendicular to the projection
of $\mathbf{B}_{0}$ on the plane of the sky, it follows that we can
write

\begin{equation}
\alpha^{2}\left(0\right)-\alpha^{2}\left(\boldsymbol{\ell}\right)=\sum_{\stackrel{{\scriptstyle i+j=1}}{i,j\geq0}}^{\infty}a_{2i,2j}\ell_{1}^{2i}\ell_{2}^{2j}.\label{eq:Taylor_ani}
\end{equation}

\noindent Accordingly, we will proceed by fitting the part within
curly braces in Equation (\ref{eq:b_alpha}) using

\begin{equation}
b^{2}\left(0\right)+\left[\alpha^{2}\left(0\right)-\alpha^{2}\left(\boldsymbol{\ell}\right)\right]=b^{2}\left(0\right)+\sum_{\stackrel{{\scriptstyle i+j=1}}{i,j\geq0}}^{\infty}a_{2i,2j}\ell_{1}^{2i}\ell_{2}^{2j}\label{eq:model}
\end{equation}

\noindent to the data for high enough values of $\ell\equiv\left|\boldsymbol{\ell}\right|$
where we expect $b^{2}\left(\boldsymbol{\ell}\right)$ to be negligible
(i.e., $b^{2}\left(\boldsymbol{\ell}\right)$ dominates at lower values
of $\ell$). We will then obtain $b^{2}\left(\boldsymbol{\ell}\right)$
by subtracting the dispersion function data (i.e., the left hand side
of Equation (\ref{eq:b_alpha})) from the aforementioned fit. This
function is the equivalent to Equation (\ref{eq:chi(ell)}) for $\chi\left(\ell\right)$
defined in Section \ref{sub:Angular-Structure-Function} for the angular
structure function analysis.

\begin{figure}
\epsscale{0.8}\plotone{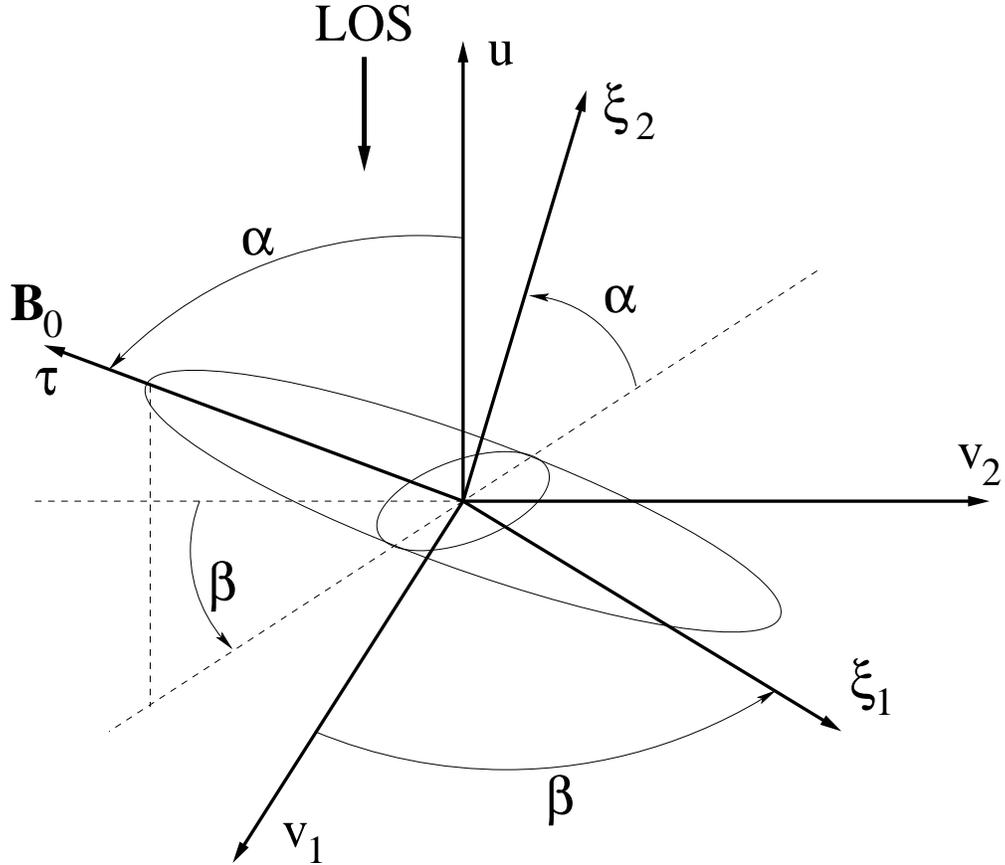}

\caption{\label{fig:ellipse}Orientation of the Gaussian ellipsoid characterizing
our idealization of magnetized turbulence in the $\left(v_{1},v_{2},u\right)$
coordinate system. The $v_{1}$ and $v_{2}$ axes define the plane
of the sky while the line of sight point along the negative $u$-axis.
The inclination angle relative to the line of sight of the ellipsoid
symmetry axis (and of $\mathbf{B}_{0}$) is given by $\alpha$ while
the angle $\beta$ defines the orientation of its projection on the
plane of the sky. }
\end{figure}

We now adopt a picture for magnetized turbulence consistent with current
models, either incompressible \citep{Goldreich1995} or compressible
\citep{Cho2003, Kowal2010}. That is, we will assume that some anisotropy
is present in the magnetized turbulent autocorrelation function where
it is expected that the correlation lengths in directions parallel
and perpendicular to the ordered magnetic field are different. Although
this is undoubtedly an idealization, we model the intrinsic autocorrelation
function of the magnetized turbulence as an ellipsoid Gaussian function,
with the symmetry axis of the ellipsoid aligned with the ordered magnetic
field $\mathbf{B}_{0}$ (see Fig. \ref{fig:ellipse}). It is possible
to analytically solve for $b^{2}\left(\boldsymbol{\ell}\right)$ for
such cases with the further assumption that the telescope beam profile
is circular Gaussian in form

\begin{equation}
H\left(\mathbf{r}\right)=\frac{1}{2\pi W^{2}}e^{-r^{2}/2W^{2}},\label{eq:beam}
\end{equation}

\noindent with $W=0.425\:\mathrm{FWHM}$ the beam radius (i.e., it
is not the beam's FWHM but its ``standard deviation'' equivalent).
A general solution for $b^{2}\left(\boldsymbol{\ell}\right)$ when
the magnetized autocorrelation function ellipsoid is at arbitrary
inclination relative to the line of sight and arbitrary projected
orientation on the plane of the sky can be derived and is given in
the Appendix. 

For M51 we will limit ourselves to three cases. One has the ellipsoid
symmetry axis (and the ordered magnetic field $\mathbf{B}_{0}$) at
an inclination angle $\alpha$ relative to the line of sight and its
projection on the plane of the sky advantageously aligned with one
of the observers coordinate axes (still on the plane of the sky; see
the Appendix). The integrated (or beam-broadened) autocorrelation
function is then analytically expressed by

\begin{equation}
b^{2}\left(\boldsymbol{\ell}\right)=\left[\frac{\left\langle B_{\mathrm{t}}^{2}\right\rangle }{N\left\langle B_{0}^{2}\right\rangle +\left\langle B_{\mathrm{t}}^{2}\right\rangle }\right]e^{-\frac{1}{2}\left[\ell_{1}^{2}/\left(\delta_{\bot}^{2}+2W^{2}\right)+\ell_{2}^{2}/\left(\mu_{2}^{2}+2W^{2}\right)\right]},\label{eq:b^2_inclined}
\end{equation}

\noindent where the number of turbulence cells contained in the gas
probed by the telescope beam is given by

\begin{equation}
N=\frac{\sqrt{\left(\delta_{\bot}^{2}+2W^{2}\right)\left(\mu_{2}^{2}+2W^{2}\right)}\,\Delta^{\prime}}{\sqrt{2\pi}\delta_{\Vert}\delta_{\bot}^{2}},\label{eq:N_inclined}
\end{equation}

\noindent with

\begin{equation}
\mu_{2}^{2}=\delta_{\Vert}^{2}\sin^{2}\left(\alpha\right)+\delta_{\bot}^{2}\cos^{2}\left(\alpha\right)\label{mu2}
\end{equation}

\noindent and $\delta_{\Vert}$ and $\delta_{\bot}$ the turbulent
correlations lengths parallel and perpendicular to the (local) ordered
magnetic field $\mathbf{B}_{0}$, respectively. The inverse of these
correlation lengths are, in effect, the corresponding widths of the
associated turbulent power spectrum (see the Equation (\ref{eq:FT})
and the associated discussion in the Appendix). We once again emphasize
that, in Equation (\ref{eq:b^2_inclined}), $\ell_{1}$ and $\ell_{2}$
are the displacements respectively perpendicular and parallel to the
projection of $\mathbf{B}_{0}$ on the plane of the sky. In Equation
(\ref{eq:N_inclined}) $\Delta^{\prime}$ is the effective depth of
the column of polarized gas probed by the telescope beam (\citealt{Houde2009};
see their Section 3.2 and Equation (45)). This elliptical Gaussian
turbulence case will actually be the last one considered in Section
\ref{sub:Two-dimensional Anisotropic}. 

The first case considered (in Section \ref{sub:Isotropic-Turbulence})
will be for isotropic turbulence when $\delta_{\Vert}=\delta_{\bot}\equiv\delta$;
Equations (\ref{eq:b^2_inclined}) and (\ref{eq:N_inclined}) then
reduce to

\begin{equation}
b^{2}\left(\ell\right)=\left[\frac{\left\langle B_{\mathrm{t}}^{2}\right\rangle }{N\left\langle B_{0}^{2}\right\rangle +\left\langle B_{\mathrm{t}}^{2}\right\rangle }\right]e^{-\ell^{2}/2\left(\delta^{2}+2W^{2}\right)}\label{eq:b^2(ell)}
\end{equation}

\noindent and

\begin{equation}
N=\frac{\left(\delta^{2}+2W^{2}\right)\Delta^{\prime}}{\sqrt{2\pi}\delta^{3}}.\label{eq:N_isotropic}
\end{equation}

\noindent Equation (\ref{eq:Taylor_ani}) is furthermore simplified
to

\begin{equation}
\alpha^{2}\left(0\right)-\alpha^{2}\left(\ell\right)=\sum_{j=1}^{\infty}a_{2j}\ell^{2j}.\label{Taylor_iso}
\end{equation}

\noindent This isotropic Gaussian turbulence is the model previously
solved and used in \citet{Houde2009}.

Also, in Section \ref{sub:hybrid} we consider what could be termed
as an ``hybrid'' model of anisotropic turbulence where Equations
(\ref{eq:b^2(ell)}) - (\ref{Taylor_iso}) for isotropic turbulence
are applied to independent analyses on displacement vectors that are
grouped in sets oriented approximately parallel or perpendicular to
the local mean magnetic field. Differences in the width of the two
turbulent autocorrelation functions can then reveal anisotropy in
the magnetized turbulence.

For all three cases, the parameters measurable by fitting the integrated
turbulent autocorrelation data with Equation (\ref{eq:b^2_inclined})
(or Equation (\ref{eq:b^2(ell)})) are the correlation lengths $\delta_{\Vert}$
and $\delta_{\bot}$ (or simply $\delta$), and the intrinsic turbulent-to-ordered
magnetic energy ratio $\left\langle B_{\mathrm{t}}^{2}\right\rangle /\left\langle B_{0}^{2}\right\rangle $
(we assume that $\Delta^{\prime}$ and the inclination angle $\alpha$
are known a priori).

\section{Observations\label{sec:Observations}}

In this paper we use the high resolution radio polarization observations
of \citet{Fletcher2011}. M51 was observed with the VLA at $\lambda6.2$
cm using the C- and D-array configurations. Standard data reduction
and imaging was carried out using AIPS to produce maps of the Stokes
parameters $I$, $Q$, and $U$. These maps were merged with an Effelsberg
100-m telescope map at the same wavelength in order to correct for
missing large-scale emission in the VLA data. The polarization angles
we are using in this paper were calculated as $\Phi=\frac{1}{2}\arctan\left(U/Q\right)$
and the polarized intensity as $P=\sqrt{Q^{2}+U^{2}}$, with a first-order
correction for the positive bias in polarization due to noise. This
is accomplished by simply subtracting the polarization uncertainty
$\sigma_{P}$ to the measured polarization intensity $P_{M}$ such
that $P^{2}\simeq P_{M}^{2}-\sigma_{P}^{2}$, which is a good approximation
when $P\geq3\sigma_{P}$ \citep{Wardle1974}. We only use the highest
resolution maps, with a FWHM of $4\arcsec$, a grid sampling of 1\arcsec,
and RMS noise of $\sigma_{I}=15\,\mu\mathrm{Jy/beam}^{-1}$ in Stokes
$I$ and $\sigma_{P}=10\,\mu\mathrm{Jy/beam}^{-1}$ in $P$. At the
assumed distance of $7.6$ Mpc for M51 \citep{Ciardullo2002} $4\arcsec\approx150$
pc. The three regions shown in Figure \ref{fig:m51_polflux} contain
520, 229, and 301 data points that verify $P\geq3\sigma_{P}$, which
are used for the corresponding analyses for the spiral arms in the
northeast and southwest, and the galaxy center, respectively.

\subsection{Data Analysis}

As stated in Sections \ref{sub:Angular-Structure-Function} and \ref{sub:Angular-Dispersion-Function},
the polarization angle $\Phi$ is the basic observable needed for
our analysis. Following the detailed discussion presented in Appendix
B of \citet{Houde2009}, given the angle difference between a pair
of data points separated by a distance $\ell_{ij}\equiv\vert\mathbf{r}_{i}-\mathbf{r}_{j}\vert$

\begin{equation}
\Delta\Phi_{ij}=\Phi_{i}-\Phi_{j}\label{eq:diff_ij}
\end{equation}

\noindent we calculate the (average) function $\left\langle \cos\left(\Delta\Phi_{ij}\right)\right\rangle _{k}$
from the data for all $\left(\ell_{k}-\Delta\ell/2\right)\leq\ell_{ij}<\left(\ell_{k}+\Delta\ell/2\right)$,
with $\ell_{k}=k\Delta\ell$ corresponding to an integer multiple
of the grid spacing $\Delta\ell=1\arcsec$. This function is then
corrected for measurement uncertainties according to

\begin{equation}
\left\langle \cos\left(\Delta\Phi_{ij}\right)\right\rangle _{k,0}\simeq\frac{\left\langle \cos\left(\Delta\Phi_{ij}\right)\right\rangle _{k}}{1-\frac{1}{2}\left\langle \sigma^{2}(\Delta\Phi_{ij})\right\rangle _{k}},\label{eq:cos_ij_corr}
\end{equation}

\noindent where the uncertainty on $\Delta\Phi_{ij}$ is given by

\begin{equation}
\sigma^{2}(\Delta\Phi_{ij})\simeq\sigma^{2}(\Phi_{i})+\sigma^{2}(\Phi_{j})-2\sigma(\Phi_{i})\sigma(\Phi_{j})e^{-\ell_{ij}^{2}/4W^{2}}\label{eq:sigma2_dphi}
\end{equation}

\noindent and $\sigma^{2}(\Phi_{i})$ is the uncertainty on $\Phi_{i}$.
Finally, the measurement uncertainties for the adopted dispersion
function $1-\left\langle \cos\left(\Delta\Phi_{ij}\right)\right\rangle _{k,0}$
is determined with 

\begin{equation}
\sigma^{2}\left[\left\langle \cos\left(\Delta\Phi_{ij}\right)\right\rangle _{k,0}\right]=\left\langle \sin\left(\Delta\Phi_{ij}\right)\right\rangle _{k}^{2}\left\langle \sigma^{2}(\Delta\Phi_{ij})\right\rangle _{k}+\frac{3}{4}\left\langle \cos\left(\Delta\Phi_{ij}\right)\right\rangle _{k}^{2}\left\langle \sigma^{4}(\Delta\Phi_{ij})\right\rangle _{k},\label{eq:sigma2(cos)}
\end{equation}

\noindent for all $\left(\ell_{k}-\Delta\ell/2\right)\leq\ell_{ij}<\left(\ell_{k}+\Delta\ell/2\right)$.
The data and results presented in the figures and tables that follow
are all based on these equations.

\subsection{Faraday rotation}

Our analysis assumes that the observed polarization angles at $\lambda6$
cm trace the orientation of the local magnetic field (we ignore the
$\pi/2$ difference between the linear polarization plane of the observed
electric field and the orientation of the magnetic field at the source
of the emission). Faraday rotation can add an extra level of complexity
to the distribution of angles, so here we estimate its contribution
to the observed angles. 

Faraday rotation can produce a systematic variation of $\Phi$ with
position due to the presence of a mean magnetic field; if the mean
field lies in the same plane as the galaxy disc then the positional
variation will occur due to the inclination of the galaxy to the line
of sight. \citet{Fletcher2011} modeled the mean magnetic field in
M51 and found that it does lie in the galaxy plane and is weak. The
\emph{maximum} observed rotation measure due to the mean field of
their model is $\mathrm{RM}\approx10\mathrm{\, rad\, m}^{2}$, which
rotates $\lambda6$ cm emission by $\Delta\Phi\approx2\deg$.

Random fluctuations of Faraday rotation, $\sigma_{\mathrm{RM}}$,
will also produce fluctuations in $\Phi$. \citet{Fletcher2011} estimated
that the intrinsic standard deviation of $\mathrm{RM}$ in M51 at
$15\arcsec$ resolution is $\sigma_{\mathrm{RM}}\approx10\mathrm{\, rad\, m}^{2}$.
At the $4\arcsec$ resolution we are using $\sigma_{\mathrm{RM}}$
will be stronger, scaling as the ratio of the beam widths, so in our
data $\sigma_{\mathrm{RM}}\approx40\,\mathrm{rad\, m}^{2}$ corresponding
to a rotation of $\Delta\Phi\approx8\deg$ at $\lambda6$ cm.

Thus Faraday rotation produces uncertainty in our dispersion functions
of $1-\langle\cos(\Delta\Phi)\rangle\approx0.01$. This uncertainty
is about an order of magnitude below the difference between the fits
using Equation (\ref{eq:Taylor_ani}) (or (\ref{Taylor_iso})) and
the observations, until decorrelation of the angles occurs (e.g.,
see the middle panels of Figs. \ref{fig:m51_iso_northeast} to \ref{fig:m51_iso_southwest}).
Note that decorrelation, i.e., where the autocorrelation function
becomes zero, mostly occurs when the dispersion function is about
$0.1$, which corresponds to an angle difference of about $30\deg$.
Therefore we will ignore the contribution of Faraday rotation to the
polarization angles in our data, other than as a source of error.
We also note that $\lambda3$ cm data also presented in \citet{Fletcher2011},
which suffer less from Faraday rotation, were not used for this analysis
because of lower spatial resolution.

\section{Results\label{sec:Results}}

In this section we present the results of the angular dispersion analyses
for isotropic and anisotropic magnetized turbulence. We used the polarization
data from the three regions identified in Figure \ref{fig:m51_polflux}:
the northeast spiral arm, the centre of the galaxy, and the southwestern
spiral arm. In all cases, we only consider measurements for which
$p\geq3\sigma_{p}$, where $p=P/I$ and $\sigma_{p}$ are the polarization
level and its uncertainty, respectively.

\subsection{Isotropic Turbulence\label{sub:Isotropic-Turbulence}}

We first consider the case of isotropic turbulence and model our data
for the three suitable regions in M51 with Equations (\ref{eq:b^2(ell)})-(\ref{Taylor_iso}).
All the pertinent functions are therefore assumed to possess an azimuthal
symmetry about the $\ell=0$ axis.

\begin{figure}
\epsscale{1.0}\plotone{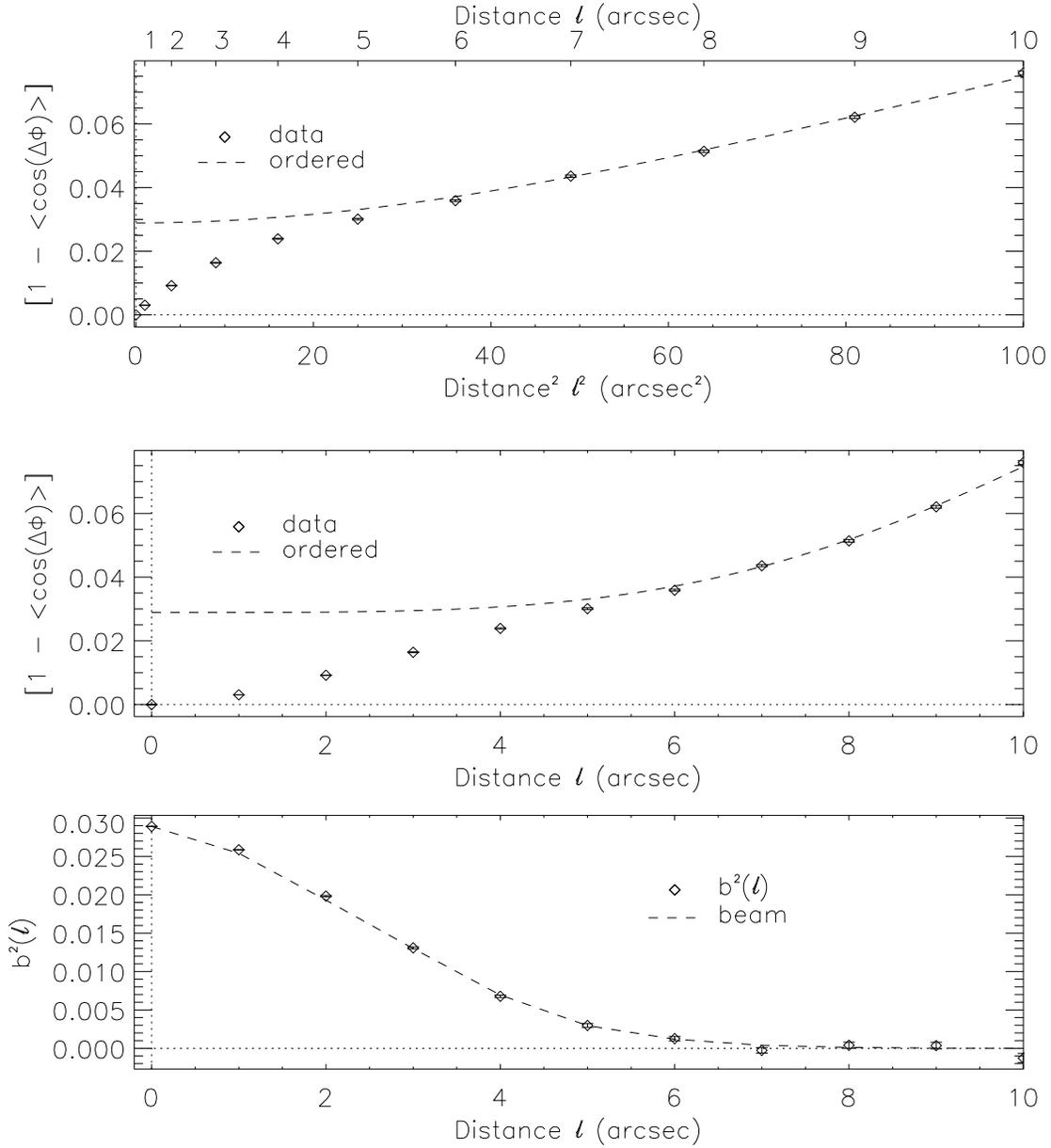}

\caption{\label{fig:m51_iso_northeast}Isotropic dispersion function for the
northeast spiral arm as a function of $\ell^{2}$ (top) $\ell$ and
(middle). The broken curve (\textquotedblleft{}ordered\textquotedblright{})
is the least-squares fit for the sum of the turbulent to total magnetic
energy ratio and the ordered component to data contained within $6\leq\ell\leq10$;
data are represented with symbols. \emph{Bottom:} the magnetized turbulence
autocorrelation function $b^{2}\left(\ell\right)$ obtained by subtracting
the data from the aforementioned fit of the middle graph. The broken
curve shows the radial profile of the \textquotedblleft{}mean autocorrelated
synthesized beam.\textquotedblright{}}
\end{figure}

Figure \ref{fig:m51_iso_northeast} shows the result of the isotropic
dispersion analysis for the northeast spiral arm as a function of
$\ell^{2}$ (top) and $\ell$ (middle). The broken curve (\textquotedblleft{}ordered\textquotedblright{})
is the least-squares fit for the sum of the integrated turbulent-to-total
magnetic energy ratio ($b^{2}\left(0\right)$ in Equation (\ref{eq:b^2(ell)}))
and the ordered component of Equation (\ref{Taylor_iso}) (i.e., Equation
(\ref{eq:model}) while using Equation (\ref{Taylor_iso}) on the
right-hand side) to the data contained within $6\leq\ell\leq10$;
data points are shown with symbols. The integrated magnetized turbulence
autocorrelation function $b^{2}\left(\ell\right)$, obtained by subtracting
the data from the aforementioned fit of the middle graph, is shown
at the bottom. The broken curve on the bottom graph shows the radial
profile of the \textquotedblleft{}autocorrelated synthesized beam.\textquotedblright{}
This represents the contribution of the synthesized beam to the (width
of) $b^{2}\left(\ell\right)$. That is, this is what $b^{2}\left(\ell\right)$
would look like in the limit where the intrinsic turbulent correlation
length $\delta$ were zero in the exponent of Equation (\ref{eq:b^2(ell)})
(i.e., disregarding its effect on the amplitude of $b^{2}\left(\ell\right)$
through $N$). It follows from this and the fact that the data points
for $b^{2}\left(\ell\right)$ fall practically on top of the autocorrelated
beam that the correlation length $\delta$ in this region of M51 is
significantly smaller than the beam size $W\simeq1\farcs70\simeq63$
pc. We also find from these graphs that $b^{2}\left(0\right)\simeq0.028$,
however, we cannot proceed any further in view of the impossibility
of determining $\delta$ for this data set.

\begin{figure}
\plotone{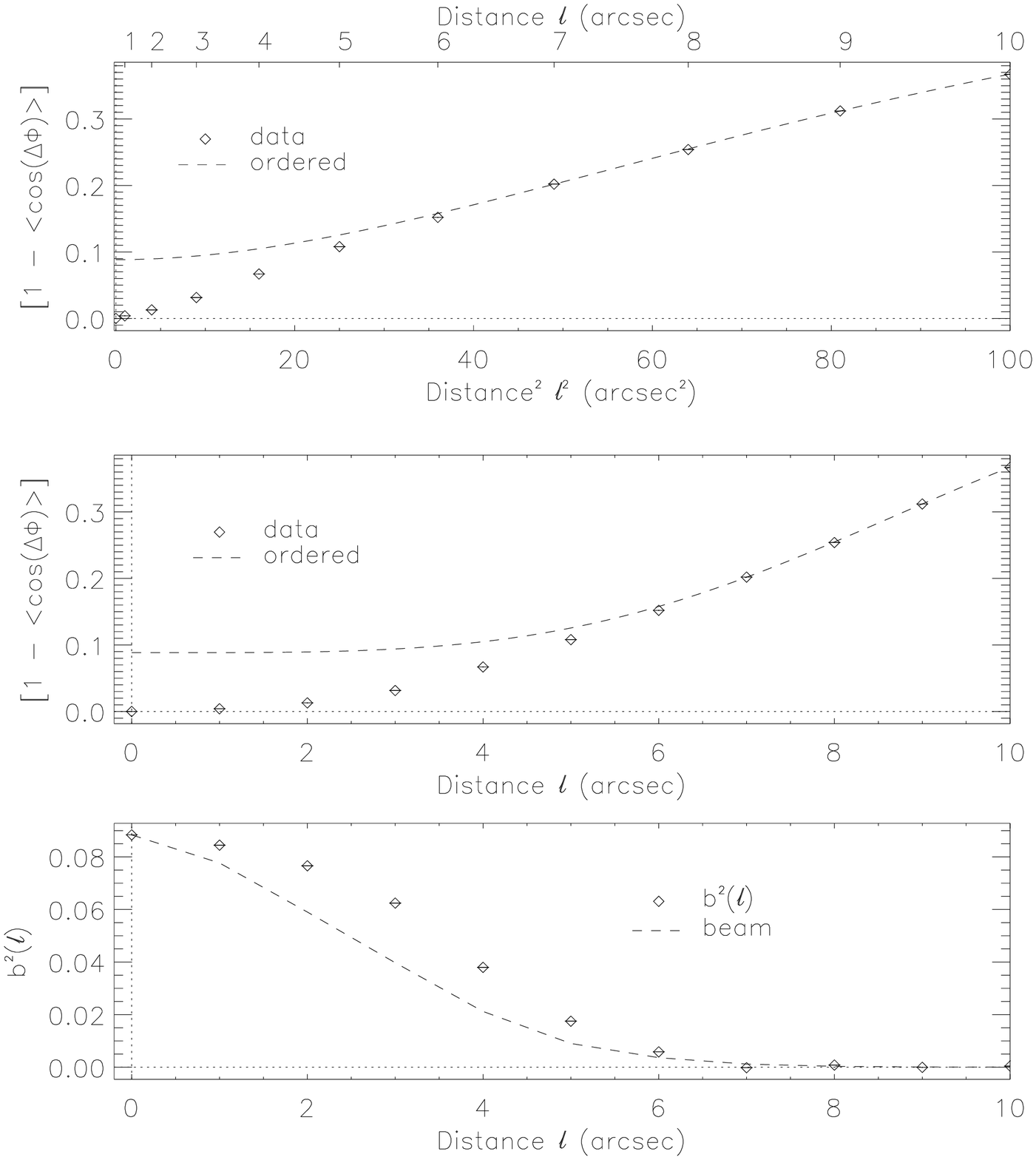}

\caption{\label{fig:m51_iso_center}Same as Figure \ref{fig:m51_iso_northeast}
but for the centre of M51.}
\end{figure}

\begin{figure}
\plotone{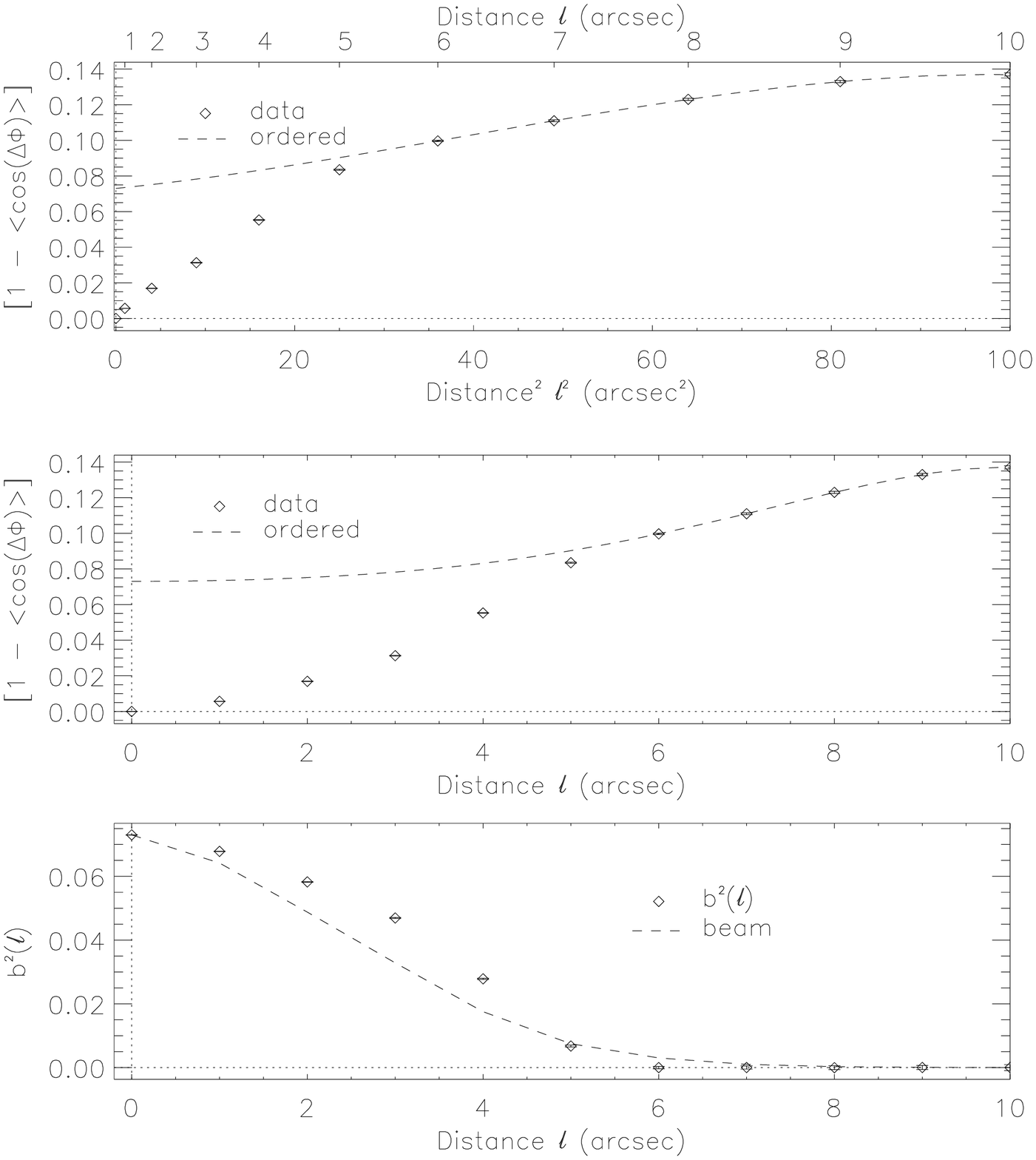}

\caption{\label{fig:m51_iso_southwest}Same as Figure \ref{fig:m51_iso_northeast}
but for the southwest spiral arm of M51.}
\end{figure}

Figures \ref{fig:m51_iso_center} and \ref{fig:m51_iso_southwest}
show the results of the dispersion analyses for the centre and the
southwest spiral arm of M51, respectively. In both cases we can clearly
see a broadening of the integrated magnetized turbulence autocorrelation
function beyond that due to the telescope beam (bottom graphs); this
is an imprint of the turbulent correlation length intrinsic to the
magnetized turbulence. Least-squares-fitting a Gaussian function to
these reveals that $\delta=2\farcs0$ (74 pc) and $1\farcs7$ (62
pc), respectively. It therefore follows that we can provide estimates
for the number of turbulent cells probed by the telescope beam and
the intrinsic turbulent-to-ordered magnetic energy ratio for these
two regions. The results are presented in Table \ref{tab:isotropic},
where we set $\Delta^{\prime}=800$ pc from \citet{Fletcher2011}.
We thus find that our results are in good agreement with those of
\citet{Fletcher2011} who estimated $B_{\mathrm{t}}/B_{0}\simeq1$
in the neighborhood of the spiral arms using a rotation measure dispersion
analysis. Similar values have been reported in previous analyses for
other sources (e.g., see \citealt{Beck1999} for NGC 6964). As will
be discussed in Section \ref{sec:Discussion}, their value of that
$2\delta\approx50$ pc is consistent to ours given the uncertainty
in some of the parameters that enter the analysis. 

\begin{deluxetable}{lccc}

\tablewidth{0pt}
\tablecolumns{4}

\tablehead{
\colhead{} & \colhead{Northeast Arm} & \colhead{Centre} & \colhead{Southwest Arm}
}

\tablecaption{Results for isotropic turbulence.
\label{tab:isotropic}}

\startdata

$\delta$ (pc)\tablenotemark{a} & \nodata & $67\pm7$ & $66\pm8$ \\
$N$\tablenotemark{b} & \nodata & $13\pm3$ & $14\pm4$ \\
$\left\langle\overline{B}_{\mathrm{t}}^{2}\right\rangle/\left\langle\overline{B}^{2}\right\rangle$\tablenotemark{c} & $0.028\pm0.002$ & $0.088\pm0.026$ & $0.072\pm0.025$ \\
$\left\langle B_{\mathrm{t}}^{2}\right\rangle/\left\langle B_{0}^{2}\right\rangle$\tablenotemark{d} & \nodata & $1.28\pm0.29$ & $1.08\pm0.29$ \\
$B_{\mathrm{t}}/B_{0}$\tablenotemark{e} & \nodata & $1.13\pm0.13$ & $1.04\pm0.14$

\enddata

\tablenotetext{a}{Turbulent correlation length ($1\arcsec=37$ pc); from the fit of Equation (\ref{eq:b^2(ell)}) to the data.}
\tablenotetext{b}{Number of turbulent cells probed by the telescope beam, using $\Delta^{\prime}=800$ pc; from Equation (\ref{eq:N_isotropic}).}
\tablenotetext{c}{Measured value for the integrated turbulent to total magnetic energy ratio, corresponding to  $b^2\left(\ell=0\right)=\left\langle B_{\mathrm{t}}^{2}\right\rangle/\left[N\left\langle B_{0}^{2}\right\rangle +\left\langle B_{\mathrm{t}}^{2}\right\rangle\right]$ (see Equation (\ref{eq:b^2(ell)})).}
\tablenotetext{d}{Turbulent to ordered magnetic energy ratio, corrected for signal integration; from the fit of Equation (\ref{eq:b^2(ell)}) to the data.}
\tablenotetext{e}{Calculated from the root of $\left\langle B_{\mathrm{t}}^{2}\right\rangle/\left\langle B_{0}^{2}\right\rangle$.}

\end{deluxetable}

\subsection{Anisotropic Turbulence\label{sub:Anisotropic-Turbulence}}

\subsubsection{``Hybrid'' Model of Anisotropic Turbulence\label{sub:hybrid}}

\begin{figure}
\epsscale{0.6}\plotone{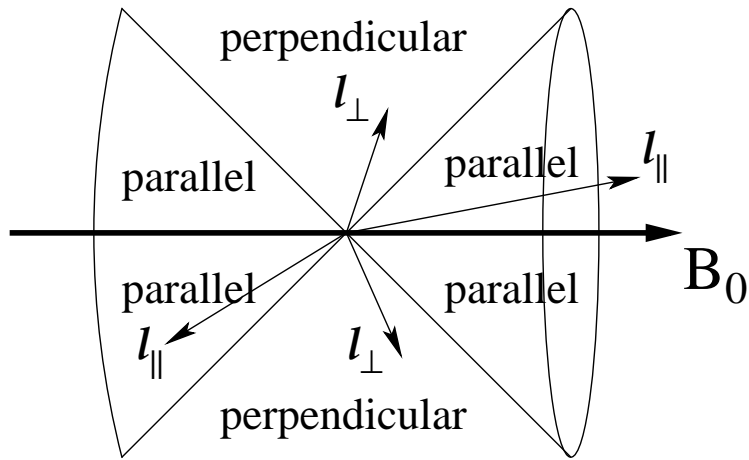}

\caption{\label{fig:m51_ani} Grouping of differences $\Delta\Phi\left(\boldsymbol{\ell}\right)$
into two sets depending whether or not $\boldsymbol{\ell}$ is oriented
in a cone whose boundaries are within $\pm45^{\circ}$ of the orientation
of the local mean magnetic field. The displacement vectors are labeled
with $\boldsymbol{\ell}_{\bot}$ or $\boldsymbol{\ell}_{\Vert}$ depending
on the case (i.e., perpendicular or parallel to the field, respectively). }
\end{figure}

We now abandon the isotropy assumption and we make a first attempt
at treating the more general case of anisotropic turbulence. To do
so, we define two separate bins of data where the polarization angle
differences used in Equation (\ref{eq:cos}) are such that the displacement
$\boldsymbol{\ell}$ is either oriented within $\pm45^{\circ}$ of
the (plane of the sky component of the) local mean magnetic field
(and labeled $\boldsymbol{\ell}_{\Vert}$) or in a direction within
$\pm45^{\circ}$ from the axis normal to it (labeled $\boldsymbol{\ell}_{\bot}$).
This is illustrated in Figure \ref{fig:m51_ani}. The orientation
of the mean field at a given position is simply approximated by averaging
polarization angles contained within a radius of $2\arcsec$. We then
perform two separate dispersion analyses that will allow us to measure
differences in the magnetized turbulence correlation lengths parallel
and perpendicular to the local field, $\delta_{\Vert}$ and $\delta_{\bot}$,
respectively. Although this analysis is not based on the more rigorous
model given in Equations (\ref{eq:b^2_inclined}) and (\ref{eq:N_inclined}),
it will allow us to look for direct evidence of anisotropic turbulence
in the same three regions of M51, as was done in the previous subsection.
A more rigorous analysis applied to M51 as a whole will follow in
Section \ref{sub:Two-dimensional Anisotropic}. 

\begin{figure}
\epsscale{1.0}\plotone{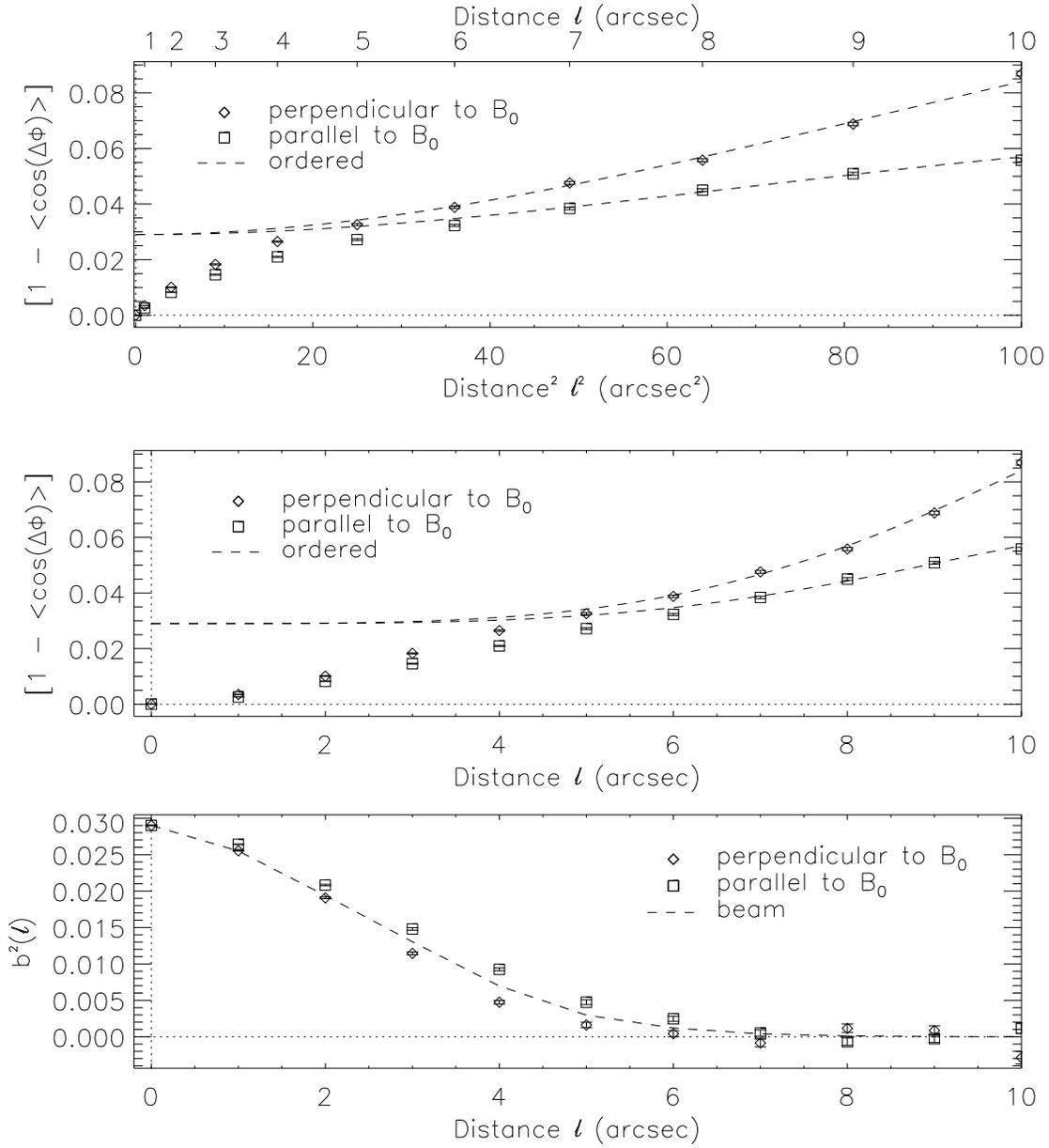}

\caption{\label{fig:m51_ani_northeast}Same as Figure \ref{fig:m51_iso_northeast}
for the northeast spiral arm, but for directions parallel and perpendicular
to the local mean magnetic field.}
\end{figure}

Figure \ref{fig:m51_ani_northeast} shows the results for the northeast
spiral arm previously analyzed under the isotropy assumption in Figure
\ref{fig:m51_iso_northeast}. For such analysis the dispersions function
parallel and perpendicular to the magnetic field must be treated simultaneously.
That is, the least-squares fits for the sum of the integrated turbulent-to-total
magnetic energy ratio and the ordered component (broken curves in
the top two graphs) to the data contained within $6\leq\ell\leq10$
are not independent since they must meet at $\ell_{\Vert}=\ell_{\bot}=0$.
These fits are thus performed simultaneously. The bottom graph shows
the integrated magnetized turbulence autocorrelation functions parallel
and perpendicular to the mean field as well as the mean autocorrelated
telescope beam, as before. Although we see evidence for anisotropy
from the separation of the two autocorrelation functions, we find
that the perpendicular function has a width that is narrower than
the contribution of the telescope beam, which is impossible. This
is most likely due to the fact that our fits (on the top two graphs)
are made with data points that are located at too low values for the
displacements $\ell_{\Vert}$ and $\ell_{\bot}$ and therefore to
some extent fail to cleanly separate the ordered and turbulent dispersion
functions (at the expense of the latter; see Sec. \ref{sec:Discussion}).
At any rate, we can infer from this analysis that $\left\langle \overline{B}_{\mathrm{t}}^{2}\right\rangle /\left\langle \overline{B}_{0}^{2}\right\rangle \simeq0.03$
(in agreement with the isotropic analysis) and that the intrinsic
magnetized turbulence autocorrelation appears to be broader along
the local magnetic field orientation than perpendicular to it.

\begin{figure}
\plotone{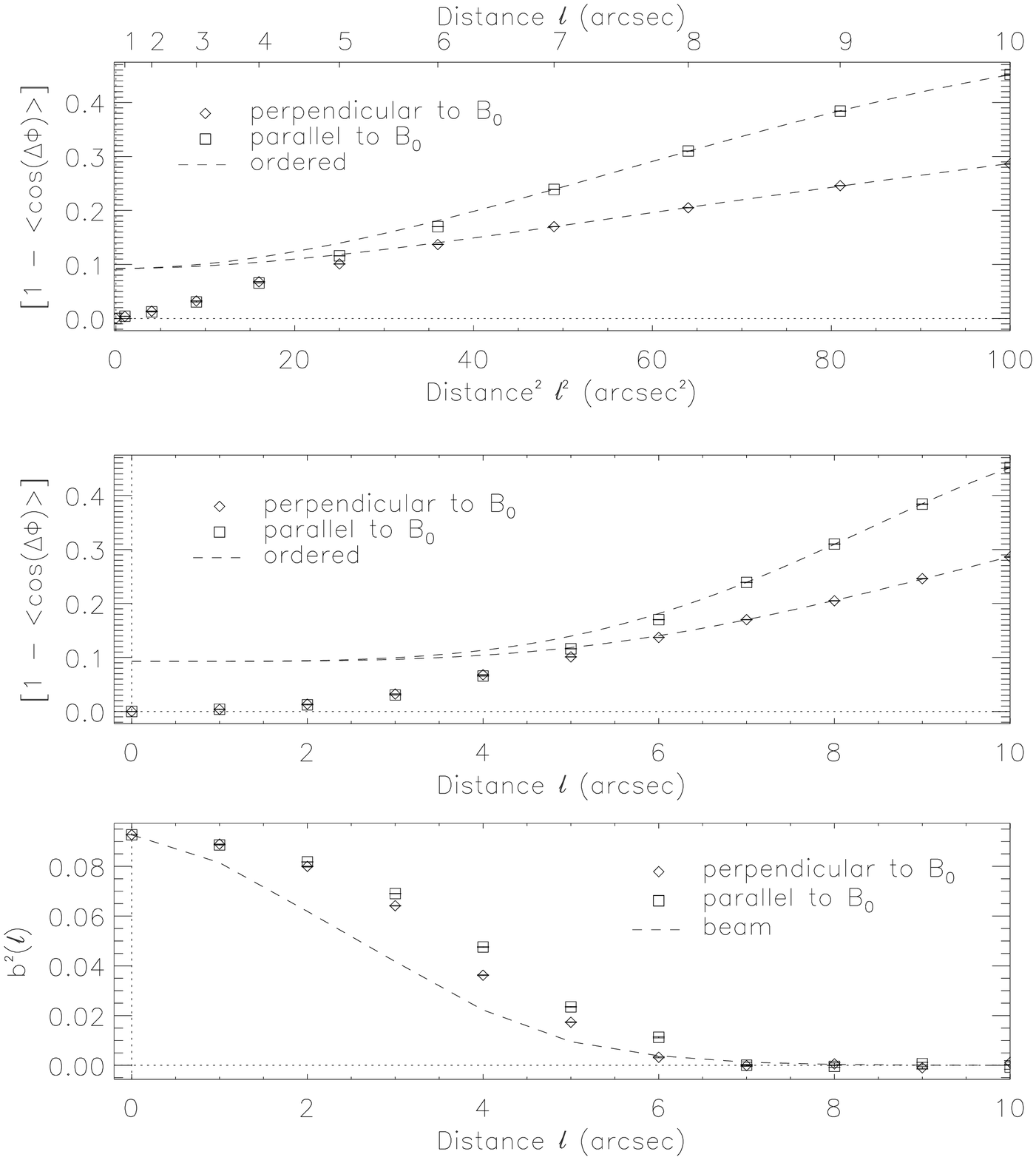}

\caption{\label{fig:m51_ani_center}Same as Figure \ref{fig:m51_ani_northeast}
but for the center of M51.}
\end{figure}

\begin{figure}
\plotone{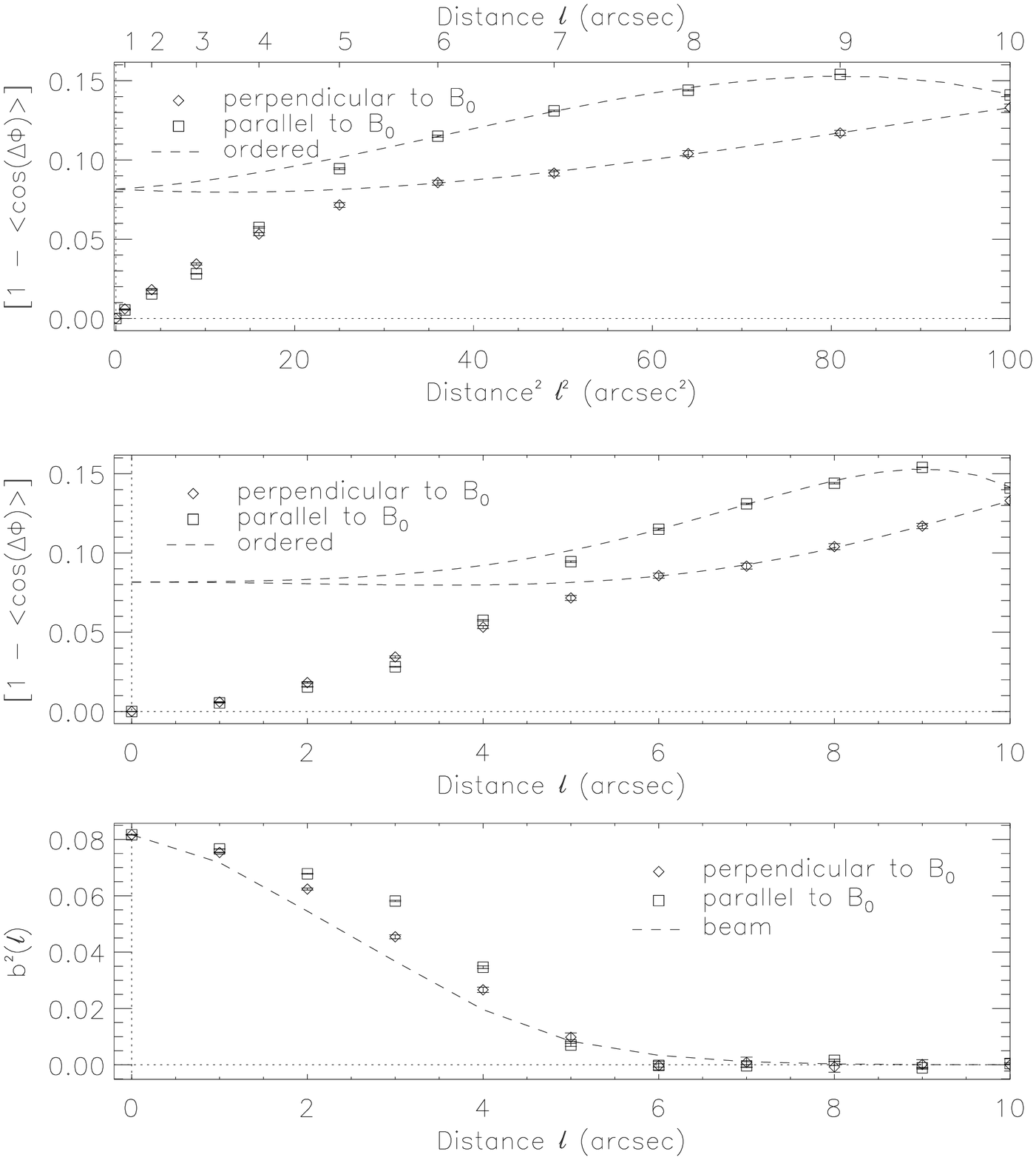}

\caption{\label{fig:m51_ani_southwest}Same as Figure \ref{fig:m51_ani_northeast}
but for the southwest spiral arm. }
\end{figure}

\begin{deluxetable}{lccc}

\tablewidth{0pt}
\tablecolumns{4}

\tablehead{
\colhead{} & \colhead{Northeast Arm} & \colhead{Centre} & \colhead{Southwest Arm}
}

\tablecaption{Results for anisotropic turbulence.
\label{tab:anisotropic}}

\startdata

$\delta_\parallel$ (pc)\tablenotemark{a} & $25\pm9$ & $111\pm7$ & $61\pm7$ \\
$\delta_\perp$ (pc)\tablenotemark{b} & \nodata & $87\pm8$ & $41\pm5$ \\
$\delta_\parallel / \delta_\perp$ & \nodata & $1.27\pm0.14$ & $1.48\pm0.25$ \\
$N$\tablenotemark{c} & \nodata & $7\pm1$ & $32\pm8$ \\
$\left\langle\overline{B}_{\mathrm{t}}^{2}\right\rangle/\left\langle\overline{B}^{2}\right\rangle$\tablenotemark{d} & $0.028\pm0.002$ & $0.093\pm0.003$ & $0.082\pm0.001$ \\
$\left\langle B_{\mathrm{t}}^{2}\right\rangle/\left\langle B_{0}^{2}\right\rangle$\tablenotemark{e} & \nodata & $0.68\pm0.10$ & $2.86\pm0.68$ \\
$B_{\mathrm{t}}/B_{0}$\tablenotemark{f} & \nodata & $0.83\pm0.06$ & $1.69\pm0.20$

\enddata

\tablenotetext{a}{Turbulent correlation length parallel to $\mathbf{B}_0$ ($1\arcsec=37$ pc); from the fit of Equation (\ref{eq:b^2(ell)}) to the data.}
\tablenotetext{b}{Same as note a, but perpendicular to $\mathbf{B}_0$.}
\tablenotetext{c}{Number of turbulent cells probed by the telescope beam, using $\Delta^{\prime}=800$ pc; from Equation (\ref{eq:N_inclined}).}
\tablenotetext{d}{Measured value for the integrated turbulent to total magnetic energy ratio, corresponding to  $b^2\left(\ell=0\right)=\left\langle B_{\mathrm{t}}^{2}\right\rangle/\left[N\left\langle B_{0}^{2}\right\rangle +\left\langle B_{\mathrm{t}}^{2}\right\rangle\right]$ (see Equation (\ref{eq:b^2_inclined})).}
\tablenotetext{e}{Turbulent to ordered magnetic energy ratio, corrected for signal integration; from the fit of Equation (\ref{eq:b^2(ell)}) to the data with  $\delta=\sqrt{\delta_\parallel \delta_\perp}$.}
\tablenotetext{f}{Calculated  from the root of $\left\langle B_{\mathrm{t}}^{2}\right\rangle/\left\langle B_{0}^{2}\right\rangle$.}
\end{deluxetable}

Figures \ref{fig:m51_ani_center} and \ref{fig:m51_ani_southwest}
show the results of the same anisotropic dispersion analysis for the
center and southwest spiral arm of M51, respectively. In these two
cases, however, we clearly resolve the anisotropy in the turbulence.
That is, we observe a separation in the integrated autocorrelations
functions (bottom graphs) along the directions parallel and perpendicular
to the local mean magnetic field. The former being the broader of
the two, which is also consistent with was observed for the northeast
spiral arm in Figure \ref{fig:m51_ani_northeast}. As we will discuss
in Section \ref{sec:Discussion} this result is consistent with theory
and simulations of incompressible \citep{Goldreich1995,Cho2002} and
compressible MHD turbulence \citep{Cho2003,Kowal2010}. 

The level of anisotropy in the turbulence can be gauged through the
parallel-to-perpendicular correlation length ratio $\delta_{\Vert}/\delta_{\bot}$,
which is measured to be approximately 1.3 and 1.5 for the centre and
southwest spiral arm, respectively. To estimate the number of turbulent
cells $N$ we use Equations (\ref{eq:N_inclined}) and (\ref{mu2}),
with $\alpha=\pi/2$. We once again find that the turbulent-to-ordered
magnetic field strength ratio is significant and hovers around unity
with $0.8\lesssim B_{\mathrm{t}}/B_{0}\lesssim1.7$. A summary of
the results is given in Table \ref{tab:anisotropic}.

\subsubsection{Two-dimensional, Anisotropic Gaussian Turbulence\label{sub:Two-dimensional Anisotropic}}

We finally perform one last anisotropic analysis by taking advantage
of the large number of reliable polarization measurements contained
in the complete map of M51 shown in Figure \ref{fig:m51_polflux}.
That is, we now consider the whole map at once without discriminating
between the different regions (as long as $p\geq3\sigma_{p}$). We
hope in doing so that the large number of measurements will allow
for the characterization of the intrinsic two-dimensional turbulence
autocorrelation function, using the Gaussian anisotropic model given
in Equations (\ref{eq:b^2_inclined}) and (\ref{eq:N_inclined}).
One would expect that the previously measured anisotropy, quantified
with $\delta_{\Vert}/\delta_{\bot}$, would become more pronounced
since we would do away with the cone-averages exemplified in Figure
\ref{fig:m51_ani}. 

\begin{figure}
\epsscale{0.9}\plotone{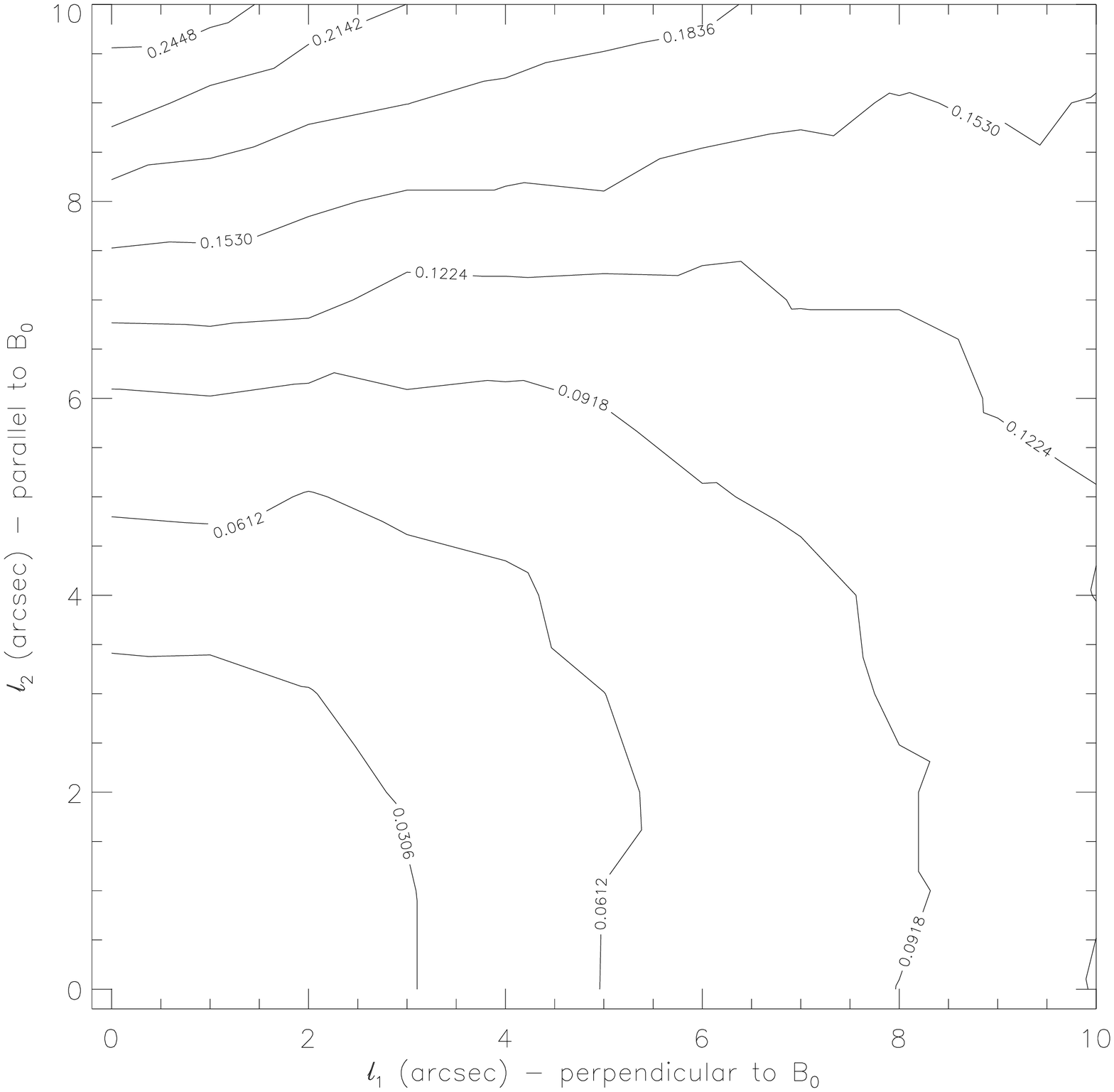}

\caption{\label{fig:m51_cont_struct}Contour plot of the two-dimensional dispersion
function for the whole M51 polarization map of \ref{fig:m51_polflux}.
This function is assumed to be even in directions parallel or perpendicular
to the mean magnetic field.}
\end{figure}

\begin{figure}
\epsscale{1.0}\plotone{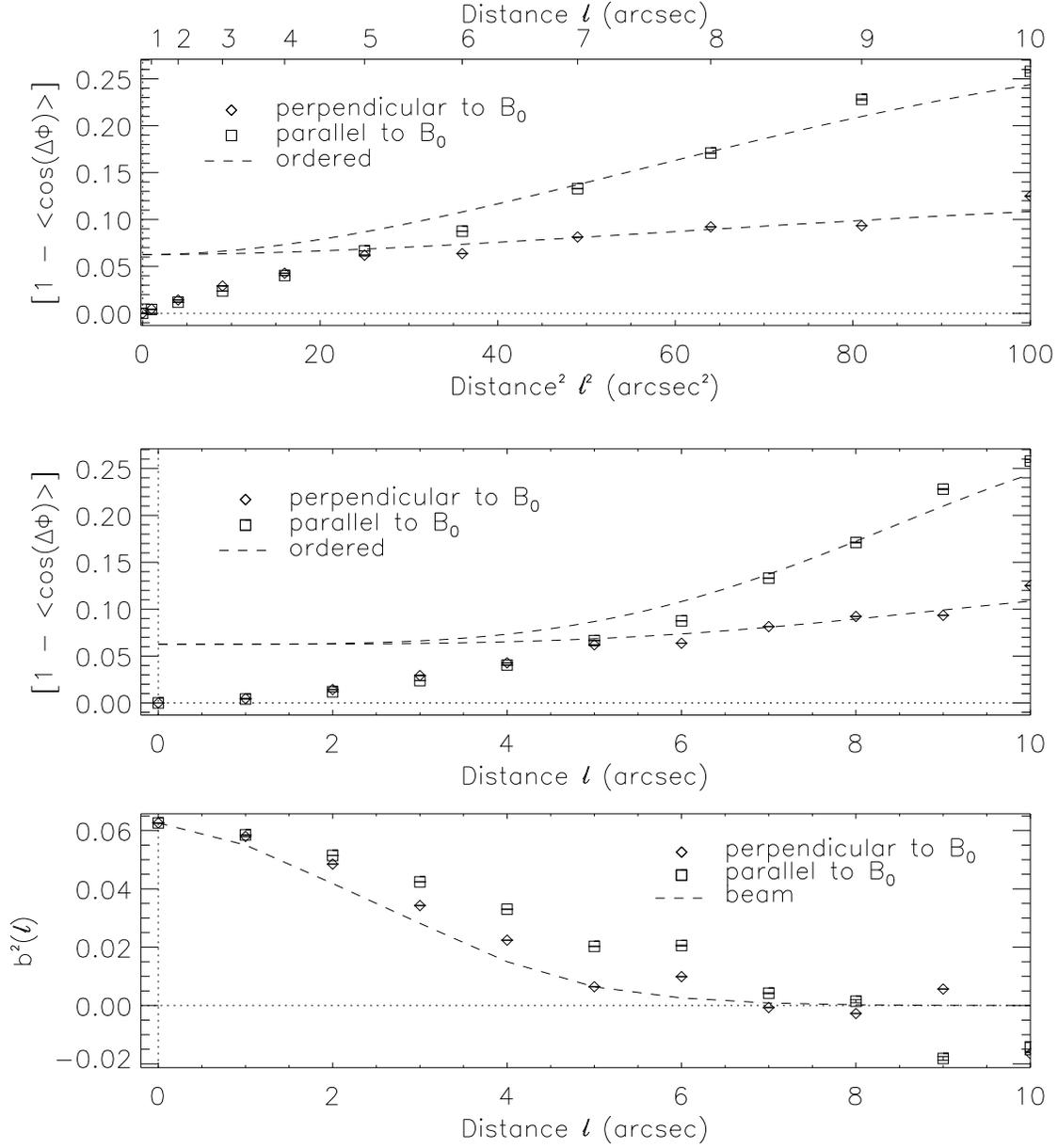}

\caption{\label{fig:m51_2d_struct}\emph{Top and middle:} cuts through the
two-dimensional dispersion function (symbols) and the ordered fits
(broken curves; using values of $7\leq\ell\leq10$) along directions
parallel and perpendicular to the mean magnetic field. \emph{Bottom:}
the corresponding profiles for the turbulence autocorrelation function.}
\end{figure}

\begin{figure}
\plotone{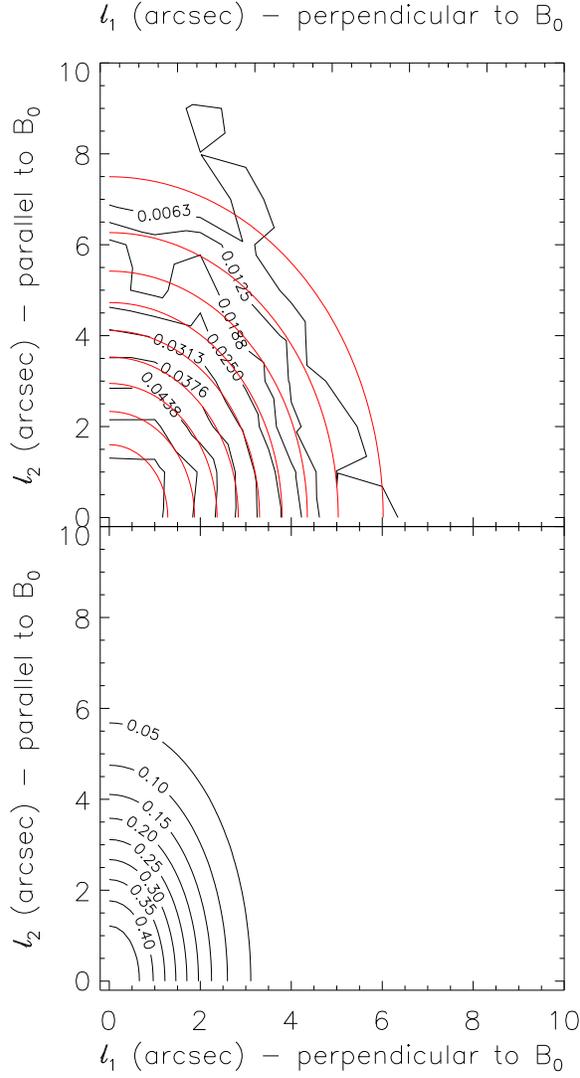}

\caption{\label{fig:m51_2d_auto}\emph{Top:} contour plot of the two-dimensional
turbulence autocorrelation function (black) and a Gaussian fit (red);
the fit is forced to be even in directions parallel and perpendicular
to the mean magnetic field. The contours are drawn at 10 to 90 percent
(10 percent increments) of the peak $b^{2}\left(0\right)=0.063$.\emph{
Bottom:} the intrinsic turbulence autocorrelation function, where
we set $W=0$, $\alpha=\pi/2$, and $N=1$ in Equations (\ref{eq:b^2_inclined})
and (\ref{mu2}), while using the values for $\delta_{\Vert}$ and
$\delta_{\bot}$ obtained with the Gaussian fit (red) in the plot
on the top graph (see Table \ref{tab:2D}). The contours are drawn
at 10 to 90 percent (10 percent increments) of the peak $\left\langle B_{\mathrm{t}}^{2}\right\rangle /\left\langle B^{2}\right\rangle =0.50$.}
\end{figure}

Figure \ref{fig:m51_cont_struct} shows the result of the two-dimensional
dispersion analysis. Only one quadrant of the contour plot of the
dispersion function is displayed since it is assumed even in directions
perpendicular and parallel to the mean magnetic field (i.e., in even
powers of $\boldsymbol{\ell}_{1}$ and $\boldsymbol{\ell}_{2}$).
Since we must now least-squares fit a two-dimensional surface corresponding
to the sum of the turbulent-to-ordered magnetic energy ratio and the
ordered component of the dispersion function (i.e., the right-hand
side of Equation (\ref{eq:b_alpha})), it is to be expected that this
fitting process will be more challenging than before. This can be
verified in Figure \ref{fig:m51_2d_struct} were cuts through the
two-dimensional dispersion and integrated turbulence autocorrelation
functions along directions parallel and perpendicular to the local
mean magnetic field are shown; data contained within $7\leq\ell\leq10$
were used to perform the aforementioned least-squares fit to the right-hand
side of Equation (\ref{eq:b_alpha}). A comparison of Figure \ref{fig:m51_2d_struct}
with any such figures stemming from the previous isotropic or anisotropic
analyses reveals that our fit to the two-dimensional dispersion function
(top and middle graphs) is unable to perfectly match the data. The
main consequence of this being the somewhat \textquotedblleft{}ragged\textquotedblright{}
appearance of the integrated two-dimensional turbulence autocorrelation
function presented in the bottom graph of Figure \ref{fig:m51_2d_struct}
(symbols). Nonetheless, it is interesting to note that we once again
find the same anisotropy as before, with the result that $\delta_{\Vert}>\delta_{\bot}$.
We sought to quantify this by performing a two-dimensional (elliptical)
Gaussian least-squares fit, using Equation (\ref{eq:b^2_inclined}),
to the integrated turbulence autocorrelation function data. This is
shown in the top panel of Figure \ref{fig:m51_2d_auto}, where a contour
plot of the aforementioned Gaussian fit (red) is superposed on that
of the integrated two-dimensional turbulence autocorrelation function
(black). For this we used the known value of $\alpha=70^{\circ}$
for M51 \citep{Tully1974}. The fit is forced to be even in directions
perpendicular and parallel to the local mean magnetic field (i.e.,
even in powers of $\boldsymbol{\ell}_{1}$ and $\boldsymbol{\ell}_{2}$;
see Equation (\ref{eq:Taylor_ani})), as the dispersion function was
also assumed to be. Although this Gaussian fit appears to be reasonable
for $\ell\leq4\arcsec$, it is not expected to be a realistic representation
since it is unlikely that the magnetized turbulence is Gaussian in
nature in M51. Nonetheless, it allows us to extract a useful approximation
to the intrinsic two-dimensional magnetized turbulence autocorrelation
function; the resulting function is shown in the bottom panel of Figure
\ref{fig:m51_2d_auto}. As stated earlier, such results are consistent
with theory and simulations of incompressible \citep{Goldreich1995,Cho2002}
and compressible MHD turbulence \citep{Cho2003,Kowal2010}. The parameters
extracted from this anisotropic analysis are also consistent with
our previous results for the three independent regions and are presented
in Table \ref{tab:2D}. 

Our results do not take into account any systematic uncertainties
on some of the parameters used to characterize M51. For example, the
effective depth $\Delta^{\prime}=800$ pc, which comes in for all
three cases treated in this section (isotropic, ``hybrid'' anisotropic,
and anisotropic turbulence), enters linearly in the evaluation of
the number of turbulent cells $N$ contained in a telescope beam (see
Equations (\ref{eq:N_inclined}) and (\ref{eq:N_isotropic})). In
turn, the relative strength of the turbulent magnetic field component
to the ordered magnetic field scales inversely with $N^{1/2}$. An
overestimation by a factor of two in $\Delta^{\prime}$, for example,
would bring a corresponding underestimate of $B_{\mathrm{t}}/B_{0}$
by $\sqrt{2}$. Furthermore, the fully anisotropic model is also dependent
on the inclination of the galaxy, which according to \citet{Tully1974}
spans $\alpha=70^{\circ}\pm5^{\circ}$. Unlike its dependency on $\Delta^{\prime}$
discussed above, the relative level of turbulence is found to be largely
insensitive to changes in $\alpha$. On the other hand the correlation
lengths are somewhat affected by such uncertainties. For example,
we find that $100\:\mathrm{pc}\geq\ell_{\Vert}\geq96\:\mathrm{pc}$
and $1.92\geq\ell_{\Vert}/\ell_{\bot}\geq1.78$ when $65^{\circ}\leq\alpha\leq75^{\circ}$.

\begin{deluxetable}{lc}

\tablewidth{0pt}
\tablecolumns{4}

\tablehead{
\colhead{} & All regions
}

\tablecaption{Results for two-dimensional anisotropic turbulence ($\alpha=70^\circ$).
\label{tab:2D}}

\startdata

$\delta_\parallel$ (pc)\tablenotemark{a} & $98\pm5$ \\
$\delta_\perp$ (pc)\tablenotemark{b} & $54\pm3$ \\
$\delta_\parallel / \delta_\perp$ & $1.83\pm0.13$ \\
$N$\tablenotemark{c} & $15\pm2$ \\
$\left\langle\overline{B}_{\mathrm{t}}^{2}\right\rangle/\left\langle\overline{B}^{2}\right\rangle$\tablenotemark{d} & $0.063\pm0.008$ \\
$\left\langle B_{\mathrm{t}}^{2}\right\rangle/\left\langle B_{0}^{2}\right\rangle$\tablenotemark{e} & $1.02\pm0.08$ \\
$B_{\mathrm{t}}/B_{0}$\tablenotemark{f} & $1.01\pm0.04$ 

\enddata

\tablenotetext{a}{Turbulent correlation length parallel to $\mathbf{B}_0$ ($1\arcsec=37$ pc); from the fit of Equation (\ref{eq:b^2_inclined}) to the data with $\alpha=70^\circ$.}
\tablenotetext{b}{Same as note a, but perpendicular to $\mathbf{B}_0$.}
\tablenotetext{c}{Number of turbulent cells probed by the telescope beam, using $\Delta^{\prime}=800$ pc; from Equation (\ref{eq:N_inclined}).}
\tablenotetext{d}{Measured integrated turbulent to total magnetic energy ratio, corresponding to  $b^2\left(\ell=0\right)=\left\langle B_{\mathrm{t}}^{2}\right\rangle/\left[N\left\langle B_{0}^{2}\right\rangle +\left\langle B_{\mathrm{t}}^{2}\right\rangle\right]$ (see Equation (\ref{eq:b^2_inclined})).}
\tablenotetext{e}{Turbulent to ordered magnetic energy ratio, corrected for signal integration; from the fit of Equation (\ref{eq:b^2_inclined}) to the data with $\alpha=70^\circ$.}
\tablenotetext{f}{Calculated  from the root of $\left\langle B_{\mathrm{t}}^{2}\right\rangle/\left\langle B_{0}^{2}\right\rangle$.}

\end{deluxetable}

\section{Discussion\label{sec:Discussion}}

Our application of the dispersion analysis of \citet{Hildebrand2009}
and \citet{Houde2009,Houde2011}, and its generalization to include
anisotropy, to M51 reveals some interesting information on the nature
of magnetized turbulence in this galaxy. As was previously mentioned,
it is important to note that both our analysis and that of \citet{Fletcher2011}
yield results that are consistent with one another, even though they
are based on completely different approaches. For example, \citet{Fletcher2011}
determined the size of a turbulent cell (i.e., approximately twice
the turbulent correlation length) by using their measured dispersion
of rotation measure, while accounting for the averaging of turbulence
inherent to the observation process (see their Equations (3) and (5)).
Their value of approximately $50$ pc for the size of a turbulent
cell can be readily compared with our estimates of $\delta\simeq65$
pc determined for isotropic turbulence in Section \ref{sub:Isotropic-Turbulence}
(see Table \ref{tab:isotropic}). It is noteworthy that both techniques
provide results that are within a factor of two or so from each other,
which is interesting considering the uncertainty in the adopted values
for some parameters (e.g., the mean electron density $\left\langle n_{\mathrm{e}}\right\rangle =0.1$
cm$^{-3}$ used in their calculations). 

\citet{Fletcher2011} were also able to discern between the contribution
of the different components of the total magnetic field. They found
that the total magnetic field $B\sim15\,\mu\mathrm{G}$ is split into
an ordered ($B_{0}$) and an isotropic (i.e., random, $B_{\mathrm{t}}$)
components each of $\sim10\,\mu\mathrm{G}$. It is important to note
that their definition of an ``ordered'' magnetic field is not restricted
to a ``mean'' field, which would result from the average of the
magnetic field vector over some suitable (large) scale. More precisely,
they define the ordered field as that which is traced by the polarized
emission. For M51 they find that the ordered magnetic field is composed
of a weak $\sim2\,\mu\mathrm{G}$ mean component and an anisotropic
random field of $\sim10\,\mu\mathrm{G}$ (we note that such a ``mean
component'' implies a field with a coherent direction, while the
anisotropic field has reversing directions). This anisotropic field,
presumably resulting from ``compression in the spiral arms or localized
enhanced shear,'' would then display a stronger azimuthal variation
and thus be responsible for most of the polarized radio emission in
M51 \citep{Fletcher2011}. Their analysis therefore yields $B_{\mathrm{t}}/B_{0}\simeq1$,
as was mentioned earlier. Since our dispersion analysis is based on
changes in the orientation of polarization vectors with length scale,
it will not be able to discriminate between anisotropic random and
mean field components of the ordered field and we cannot comment on
its detailed nature. We note, however, that our estimates $B_{\mathrm{t}}/B_{0}\simeq1.13$
and $1.04$ determined for the isotropic turbulence case in Section
\ref{sub:Isotropic-Turbulence} are in excellent agreement with that
of \citet{Fletcher2011} (see Table \ref{tab:isotropic}). The typical
degrees of polarization, at $\lambda6\:\mathrm{cm}$ and $4\arcsec$
resolution, in the three regions that we consider are $p=34\%$ in
the northeast arm, $p=29\%$ in the southwest arm, and $p=15\%$ in
the centre: in calculating these values we have assumed that $25\%$
of the continuum emission at $\lambda6\:\mathrm{cm}$ is thermal \citep{Fletcher2011}.
For the arms these data are in good agreement with the degree of polarization
expected when $B_{t}/B_{0}\approx1$, as $p=0.7\: B_{0}^{2}/(B_{0}^{2}+B_{t}^{2})\approx0.35$.
This shows that the estimates of $B_{t}/B_{0}$ derived from our polarization
angle dispersion analysis are compatible with independent methods
of interpreting polarization data. In the centre of M51 the physical
environment is different from the rest of the disc, due to the presence
of an AGN and jet, and so the fraction of thermal emission and thus
the degree of polarization are harder to estimate. In addition, in
this region the polarized emission can originate from a different
location than a large fraction of the total synchrotron emission,
again making a useful estimate of $p$ difficult.

Perhaps the most important result stemming from our analysis is the
clear detection of ani\-so\-tropy in the magnetized turbulence.
Whether we consider the three analyzable regions of M51 separately
or together we consistently find that the turbulent correlation length
is larger in a direction parallel to the mean orientation of the local
magnetic field than in a direction perpendicular to it (see Tables
\ref{tab:anisotropic} and \ref{tab:2D}). As was mentioned in Section
\ref{sub:Anisotropic-Turbulence}, this result is predicted by current
theories for incompressible and compressible MHD turbulence \citep{Goldreich1995,Cho2002,Cho2003,Kowal2010}.
Such anisotropy has also been observed in the Taurus \citep{Heyer2008}
and Orion molecular clouds \citep{Shadi2012} within our Galaxy. The
level of anisotropy we observe in M51, which we quantify with the
parallel-to-perpendicular correlation length ratio $\delta_{\Vert}/\delta_{\bot}$,
is significant and in qualitative agreement with numerical simulations
of magnetized turbulence. Our contour plot of the turbulent autocorrelation
function (see the bottom panel of Figure \ref{fig:m51_2d_auto}) can
be compared with the simulations of incompressible MHD turbulence
of \citet{Cho2002}, for example. In particular, the turbulent velocity
correlation function presented in their Figure 6 has an appearance
that is similar to our derived intrinsic two-dimensional magnetized
turbulence autocorrelation function. We would expect such similarities
between these two types of autocorrelation functions under the flux-freezing
approximation, which should hold in the medium probed with synchrotron
polarization observations. Our clearest measure of anisotropy uses
the more comprehensive Gaussian model defined with Equations (\ref{eq:b^2_inclined})
to (\ref{mu2}), where the polarization data were analyzed for M51
as a whole without discriminating between the different regions. The
significant amount of anisotropy thus measured, with $\delta_{\Vert}/\delta_{\bot}\simeq1.8$,
is a statement of the importance of magnetic fields on the dynamics
of the gas probed by the observations. We should also note that this
anisotropy of the magnetized turbulent autocorrelation function is
different from that discussed by \citet{Fletcher2011}, which pertains
to the relative intensity of the two orthogonal components of the
random magnetic field. Our analysis cannot say anything concerning
any such anisotropy in the field strength, it can only inform us on
the relative turbulent energy contained in the magnetic field through
measurements of $\left\langle B_{\mathrm{t}}^{2}\right\rangle /\left\langle B_{0}^{2}\right\rangle $,
for example.

\subsection{Shortcomings of the Dispersion Analysis}

Although the quality of the data and the high resolution with which
they were obtained allowed us to determine some fundamental parameters
that characterize magnetized turbulence in M51, we expect that a slightly
higher resolution and sampling rate would result in an even more exhaustive
analysis. As was shown by \citet{Houde2011} using submillimeter dust
polarization data for Galactic molecular clouds, spatial resolutions
resulting in smaller telescope beams such that $\delta\gtrsim\sqrt{2}W$
not only allow the determination of the same parameters uncovered
by the present analysis, they can also potentially reveal the underlying
turbulent power spectrum. This is because the beam-broadened turbulent
autocorrelation function $b^{2}\left(\boldsymbol{\ell}\right)$ is
related to the turbulent power spectrum $b^{2}\left(\mathbf{k}_{v}\right)$
through a simple Fourier transform. It is then found that 

\begin{equation}
b^{2}\left(\mathbf{k}_{v}\right)=\frac{1}{\left\langle \overline{B}^{2}\right\rangle }\left\Vert H\left(\mathbf{k}_{v}\right)\right\Vert ^{2}\left[\frac{1}{2\pi}\int\mathcal{R}_{\mathrm{t}}\left(\mathbf{k}_{v},k_{u}\right)\mathrm{sinc}^{2}\left(\frac{k_{u}\Delta}{2}\right)dk_{u}\right],\label{eq:TPS}
\end{equation}

\noindent with $\mathcal{R}_{\mathrm{t}}\left(\mathbf{k}_{v},k_{u}\right)$
and $H\left(\mathbf{k}_{v}\right)$ the Fourier transforms of the
intrinsic turbulent autocorrelation function (i.e., not beam-broadened)
and the telescope beam, respectively (see the Appendix, and \citealt{Houde2011}
for a detailed discussion). It follows that beams of smaller spatial
extent than the intrinsic turbulent autocorrelation function will
have a broader spectral coverage that will reduce their filtering
effect on the power spectrum. It then becomes possible to effectively
invert Equation (\ref{eq:TPS}) to reveal the underlying power spectrum
(through some ``deconvolution'' techniques, for example). In such
cases, it is not necessary to assume any model for the turbulence,
such as the Gaussian form used in our analysis. The measured turbulent
power spectrum could thus be modeled directly from the data and compared
to candidate theories for magnetized turbulence. 

As can be seen from our results for $b^{2}\left(\boldsymbol{\ell}\right)$
in Figures \ref{fig:m51_iso_northeast}-\ref{fig:m51_iso_southwest},
\ref{fig:m51_ani_northeast}-\ref{fig:m51_ani_southwest}, and \ref{fig:m51_2d_struct},
however, the contribution of the correlation length to the width of
the beam-broadened turbulent autocorrelation function (approximately
gauged through the ratio $\delta^{2}/\left(\delta^{2}+2W^{2}\right)$;
see Equations (\ref{eq:N_inclined}) and (\ref{eq:N_isotropic}))
is typically modest, implying that the spectral filtering of the telescope
beam is too severe to recover the intrinsic turbulent power spectrum.
But even a relatively modest increase in spatial resolution, e.g.,
by a factor of a few, could allows us to recover the power spectrum
in future observations.

Another negative impact of a larger telescope beam and its broadening
of the autocorrelation function $b^{2}\left(\boldsymbol{\ell}\right)$
is that it renders more difficult the separation of the small and
large scale components present in the dispersion function. For M51
this means that the scale of the turbulence, quantified with $\delta$,
can get ``mixed up'' with the larger scale of the spiral structure
through its artificial broadening to $\delta^{2}+2W^{2}$ caused by
the beam. As alluded to in Section \ref{sub:Isotropic-Turbulence},
this may be a reason why we were unable to see any contribution from
$\delta$ to the width of $b^{2}\left(\boldsymbol{\ell}\right)$ in
our analyses of the northeast arm (see Figs. \ref{fig:m51_iso_northeast}
and \ref{fig:m51_ani_northeast}). More precisely, we were unable
to cleanly separate the large from the small scale using our Taylor
expansions, i.e., Equations (\ref{eq:Taylor_ani}) and (\ref{Taylor_iso}).
This is probably also true, but to a lesser extent, for the other
two regions studied, as can be seen from the absence of significant
``tails'' for $\ell\gtrsim5\arcsec$ in $b^{2}\left(\boldsymbol{\ell}\right)$.
An increase in spatial resolution would resolve this issue, which
is likely to bring some error in our determination of the correlation
length scales and turbulent to total energy ratios. We expect this
error to be small, but it is not possible to quantify it at this point.

Finally, we wish to once again emphasize that the choice of a Gaussian
turbulence model is unlikely to be realistic for M51. But in view
of the aforementioned impossibility to uncover the underlying turbulent
power spectrum because of the significant spectral beam filtering,
this model, which can be solved analytically, allows us to quantify
key parameters that characterize magnetized turbulence. Furthermore,
our more comprehensive model for anisotropic Gaussian turbulence defined
with Equations (\ref{eq:b^2_inclined}) and (\ref{eq:N_inclined})
implicitly assumes that the $N$ ellipsoid turbulent cells contained
in the column of gas probed by the telescope beam have the same spatial
orientation in relation to the local magnetic field. This is clearly
unlikely to be true across the beam ($\mathrm{FWHM}\simeq148$ pc),
or through the thickness ($\sim800$ pc) and the extent of the studied
regions on the galactic disk. It is therefore more realistic to view
the correlation lengths $\delta_{\Vert}$ and $\delta_{\bot}$ as
some averages representative for magnetized turbulence in M51.

\section{Summary\label{sub:Summary}}

We conducted a dispersion analysis, using a generalization of the
technique of \citet{Houde2009} to previously published high-resolution
synchrotron polarization data \citep{Fletcher2011} with the goal
of charactering magnetized turbulence in M51. We first analyzed three
distinct regions (the center of the galaxy, and the northwest and
southwest spiral arms) and measured the turbulent correlation length
scale from the width of the magnetized turbulent correlation function
for two regions and detected the imprint of anisotropy in the turbulence
for all three. Furthermore, analyzing the galaxy as a whole allowed
us to determine a two-dimensional Gaussian model for the magnetized
turbulence in M51. We measured the turbulent correlation scales along
and perpendicular to the local mean magnetic field to be, respectively,
$\delta_{\Vert}=98\pm5$ pc and $\delta_{\bot}=54\pm3$ pc, while
the turbulent to ordered magnetic field strength ratio is found to
be $B_{\mathrm{t}}/B_{0}=1.01\pm0.04$. These results are in agreement
with those of \citet{Fletcher2011}, who performed a Faraday rotation
dispersion analysis of the same data. Finally, our detection of anisotropy,
quantified with a parallel-to-perpendicular correlation length ratio
with $\delta_{\Vert}/\delta_{\bot}\simeq1.83\pm0.13$, is consistent
with current magnetized turbulence theories.

\acknowledgements{M.H.'s research is funded through the NSERC Discovery Grant, Canada
Research Chair, and Western's Academic Development Fund programs. }

\appendix
\section{Anisotropic Gaussian Magnetized Turbulence}\label{sec:solution}

\noindent The cloud- and beam-integrated magnetic field is defined
with

\begin{equation}
\overline{\mathbf{B}}\left(\mathbf{r}\right)=\iint H\left(\mathbf{r}-\mathbf{a}\right)\left[\frac{1}{\Delta}\int_{0}^{\Delta}F_{0}\left(\mathbf{a},z\right)\mathbf{B}\left(\mathbf{a},z\right)dz\right]d^{2}a,\label{eq:Bbar}
\end{equation}

\noindent where the beam profile is denoted by $H\left(\mathbf{r}\right)$,
while the weighting function $F_{0}\left(\mathbf{a},z\right)\geq0$
is the (ordered) polarized emission associated with the magnetic field
$\mathbf{B}\left(\mathbf{a},z\right)$, and $\Delta$ is the maximum
depth of the cloud along any line of sight. The integrated autocorrelation
function is then 

\begin{equation}
\left\langle \overline{\mathbf{B}}_{\mathrm{t}}\mathbf{\cdot}\overline{\mathbf{B}}_{\mathrm{t}}\mathbf{\left(\boldsymbol{\ell}\right)}\right\rangle =\iint\iint H\left(\mathbf{a}\right)H\left(\mathbf{a}^{\prime}+\boldsymbol{\ell}\right)\left[\frac{2}{\Delta}\int_{0}^{\Delta}\left(1-\frac{u}{\Delta}\right)\mathcal{R}_{\mathrm{t}}\left(\mathbf{v},u\right)du\right]d^{2}a^{\prime}d^{2}a,\label{eq:int_auto}
\end{equation}

\noindent with $\mathcal{R}_{\mathrm{t}}\left(v,u\right)=\left\langle F_{0}\left(\mathbf{a},z\right)F_{0}\left(\mathbf{a}^{\prime},z^{\prime}\right)\right\rangle \left\langle \mathbf{B}_{\mathrm{t}}\left(\mathbf{a},z\right)\cdot\mathbf{B}_{\mathrm{t}}\left(\mathbf{a}^{\prime},z^{\prime}\right)\right\rangle $,
$u=\left|z^{\prime}-z\right|$, and $\mathbf{v}=\mathbf{a}^{\prime}-\mathbf{a}$
($\equiv v_{1}\mathbf{e}_{1}+v_{2}\mathbf{e}_{2}$; \citealt{Houde2009}).
We refer to $\left(v_{1},v_{2},u\right)$ as the observer coordinate
system, the $v_{1}$ and $v_{2}$ axes define the plane of the sky,
while the line of sight point along the negative $u$-axis.

The assumptions of statistical independence between the turbulent
and ordered magnetic fields, homogeneity in their strength across
the source, and of overall stationarity previously stated in Section
\ref{sub:Angular-Dispersion-Function} are all required to arrive
at Equations (\ref{eq:Bbar}) and (\ref{eq:int_auto}). Of these,
the assumption of homogeneity is particularly useful for analyzing
our data. This is because synchrotron polarization signals bring in
the complication that the weighting function $F_{0}\left(\mathbf{a},z\right)$
is also a function of the magnetic field strength (approximately proportional
to its second power), and would therefore appear to significantly
jeopardize any calculations stemming form Equation (\ref{eq:int_auto}).
However, this dependency is seen to disappear in the calculation of
the angular dispersion function (Equation (\ref{eq:cos})) when homogeneity
is assumed, since this weighting function will have the same proportionality
factor (due to the field strength) at all points in the source (i.e.,
in the integrands of Equations (\ref{eq:Bbar}) and (\ref{eq:int_auto})).
Our analysis can then proceed in a manner similar to the simpler case
of polarization dust emission signals, where there is no linked between
the value of $F_{0}\left(\mathbf{a},z\right)$ and the strength of
the magnetic field \citep{Houde2009}.

In cases where the autocorrelation function for anisotropic magnetized
turbulence is idealized with a Gaussian ellipsoid we write

\begin{eqnarray}
R_{\mathrm{t}}\left(\boldsymbol{\xi},\tau\right) & = & \left\langle F_{0}^{2}\left(\boldsymbol{\xi},\tau\right)\right\rangle \left\langle B_{\mathrm{t}}^{2}\right\rangle e^{-\frac{1}{2}\left(\tau^{2}/\delta_{\Vert}^{2}+\xi^{2}/\delta_{\bot}^{2}\right)}\nonumber \\
 & \simeq & \left\langle F_{0}^{2}\right\rangle \left\langle B_{\mathrm{t}}^{2}\right\rangle e^{-\frac{1}{2}\left(\tau^{2}/\delta_{\Vert}^{2}+\xi^{2}/\delta_{\bot}^{2}\right)},\label{eq:Rt(greek)}
\end{eqnarray}

\noindent where $\boldsymbol{\xi}$ ($=\left(\xi_{1},\xi_{2}\right)$)
is a two-dimensional displacement vector perpendicular to the orientation
of the local ordered magnetic field $\mathbf{B}_{0}$ and $\tau$
is the displacement along $\mathbf{B}_{0}$; a prolate example is
shown in Figure \ref{fig:ellipse} along with the relationship between
the $\left(v_{1},v_{2},u\right)$ and $\left(\xi_{1},\xi_{2},\tau\right)$
coordinate systems. We\textbf{ }adopt a model for anisotropic magnetized
turbulence where the symmetry axis of the ellipsoid is aligned with
$\mathbf{B}_{0}$; this also implies that the length scale of the
ordered field is much larger than the correlation lengths $\delta_{\Vert}$
and $\delta_{\bot}$ characterizing the turbulent field $\mathbf{B}_{\mathrm{t}}$.
The function $\left\langle F_{0}^{2}\left(\boldsymbol{\xi},\tau\right)\right\rangle $
is the autocorrelation of the ordered polarized emission, which we
approximate to a constant $\left\langle F_{0}^{2}\right\rangle \equiv\left\langle F_{0}^{2}\left(\boldsymbol{0},0\right)\right\rangle $
in Equation (\ref{eq:Rt(greek)}) as it is assumed that its correlation
length is also significantly larger than $\delta_{\Vert}$ and $\delta_{\bot}$.
We seek to express this function (i.e., Equation (\ref{eq:Rt(greek)}))
using the observer coordinates $\left(v_{1},v_{2},u\right)$. Referring
to Figure \ref{fig:ellipse}, the inclination angle relative to the
line of sight of the ellipsoid symmetry axis (and of $\mathbf{B}_{0}$)
is given by $\alpha$, while the angle $\beta$ defines the orientation
of its projection on the plane of the sky. The precise relationship
between the two coordinate systems is

\begin{eqnarray}
\xi_{1} & = & v_{1}\cos\left(\beta\right)+v_{2}\sin\left(\beta\right)\nonumber \\
\xi_{2} & = & -v_{1}\cos\left(\alpha\right)\sin\left(\beta\right)+v_{2}\cos\left(\alpha\right)\cos\left(\beta\right)+u\sin\left(\alpha\right)\label{eq:transform}\\
\tau & = & v_{1}\sin\left(\alpha\right)\sin\left(\beta\right)-v_{2}\sin\left(\alpha\right)\cos\left(\beta\right)+u\cos\left(\alpha\right),\nonumber 
\end{eqnarray}

\noindent such that

\begin{equation}
\frac{\tau^{2}}{\delta_{\Vert}^{2}}+\frac{\xi^{2}}{\delta_{\bot}^{2}}=\frac{u^{2}}{\eta^{2}}+\frac{v_{1}^{2}}{\kappa_{1}^{2}}+\frac{v_{2}^{2}}{\kappa_{2}^{2}}-\frac{2}{\kappa_{12}^{2}}v_{1}v_{2}+\frac{2}{\sigma_{1}^{2}}uv_{1}-\frac{2}{\sigma_{2}^{2}}uv_{2},\label{eq:exp_trans}
\end{equation}

\noindent with

\begin{eqnarray}
\frac{1}{\eta^{2}} & = & \frac{\cos^{2}\left(\alpha\right)}{\delta_{\Vert}^{2}}+\frac{\sin^{2}\left(\alpha\right)}{\delta_{\bot}^{2}}\nonumber \\
\frac{1}{\sigma_{1}^{2}} & = & \left(\frac{1}{\delta_{\Vert}^{2}}-\frac{1}{\delta_{\bot}^{2}}\right)\sin\left(\alpha\right)\cos\left(\alpha\right)\sin\left(\beta\right)\nonumber \\
\frac{1}{\sigma_{2}^{2}} & = & \left(\frac{1}{\delta_{\Vert}^{2}}-\frac{1}{\delta_{\bot}^{2}}\right)\sin\left(\alpha\right)\cos\left(\alpha\right)\cos\left(\beta\right)\nonumber \\
\frac{1}{\kappa_{1}^{2}} & = & \frac{\sin^{2}\left(\alpha\right)\sin^{2}\left(\beta\right)}{\delta_{\Vert}^{2}}+\frac{\cos^{2}\left(\beta\right)+\cos^{2}\left(\alpha\right)\sin^{2}\left(\beta\right)}{\delta_{\bot}^{2}}\label{eq:coef1}\\
\frac{1}{\kappa_{2}^{2}} & = & \frac{\sin^{2}\left(\alpha\right)\cos^{2}\left(\beta\right)}{\delta_{\Vert}^{2}}+\frac{\sin^{2}\left(\beta\right)+\cos^{2}\left(\alpha\right)\cos^{2}\left(\beta\right)}{\delta_{\bot}^{2}}\nonumber \\
\frac{1}{\kappa_{12}^{2}} & = & \left(\frac{1}{\delta_{\Vert}^{2}}-\frac{1}{\delta_{\bot}^{2}}\right)\sin^{2}\left(\alpha\right)\sin\left(\beta\right)\cos\left(\beta\right).\nonumber 
\end{eqnarray}

\noindent Inserting Equations (\ref{eq:exp_trans}) and (\ref{eq:coef1})
in Equation (\ref{eq:Rt(greek)}) we can express the turbulent autocorrelation
function with a dependency on $\left(v_{1},v_{2},u\right)$. 

It is advantageous to solve Equation (\ref{eq:int_auto}) by considering
its Fourier transform (i.e., the turbulent power spectrum; \citealt{Houde2009, Houde2011})

\begin{equation}
\overline{B}_{\mathrm{t}}^{2}\left(\mathbf{k}_{v}\right)=\left\Vert H\left(\mathbf{k}_{v}\right)\right\Vert ^{2}\left[\frac{1}{2\pi}\int\mathcal{R}_{\mathrm{t}}\left(\mathbf{k}_{v},k_{u}\right)\mathrm{sinc}^{2}\left(\frac{k_{u}\Delta}{2}\right)dk_{u}\right],\label{eq:FT}
\end{equation}

\noindent with the correspondence $\left(\mathbf{v},u\right)\rightleftharpoons\left(\mathbf{k}_{v},k_{u}\right)$
between the two domains, and then recover the autocorrelation function
through the inverse Fourier transform

\begin{equation}
\left\langle \overline{\mathbf{B}}_{\mathrm{t}}\mathbf{\cdot}\overline{\mathbf{B}}_{\mathrm{t}}\mathbf{\left(\boldsymbol{\ell}\right)}\right\rangle =\frac{1}{\left(2\pi\right)^{2}}\iint\overline{B}_{\mathrm{t}}^{2}\left(\mathbf{k}_{v}\right)e^{i\mathbf{k}_{v}\cdot\boldsymbol{\ell}}d^{2}k_{v}.\label{eq:IFT}
\end{equation}

\noindent We note that the two-dimensional power spectrum given by
Equation (\ref{eq:FT}) could be compared to a Kolmogorov-type spectrum,
for example, by first multiplying it by $2\pi k_{v}$ (for a three-dimensional
spectrum a factor of $4\pi k_{v}^{2}$ would be required; see \citealt{Houde2009,Houde2011}).

The solution for this problem mostly rests on the repeated application
of the following relation for the integration of Gaussian functions

\begin{equation}
\int_{-\infty}^{\infty}e^{-\left(a^{2}x^{2}+bx\right)}dx=\frac{\sqrt{\pi}}{a}e^{\left(b/2a\right)^{2}},\label{eq:Gauss_int}
\end{equation}

\noindent with $a$ and $b$ some constants. Considering Equation
(\ref{eq:Gauss_int}) and the fact that the Fourier transform of a
rotated function equals the rotated version of the Fourier transform
of the unrotated function (i.e., when $\alpha=\beta=0$; see Appendix
4 of \citealt{Houde2007}), the Fourier transform of the turbulent
autocorrelation function can be calculated to be

\begin{equation}
\mathcal{R}_{\mathrm{t}}\left(\mathbf{k}_{v},k_{u}\right)=\left\langle F_{0}^{2}\right\rangle \left\langle B_{\mathrm{t}}^{2}\right\rangle \left(2\pi\right)^{3/2}\delta_{\Vert}\delta_{\bot}^{2}e^{-\frac{1}{2}\left(\epsilon^{2}k_{u}^{2}+\mu_{1}^{2}k_{1}^{2}+\mu_{2}^{2}k_{2}^{2}-2\mu_{12}^{2}k_{1}k_{2}+2\gamma_{1}^{2}k_{u}k_{1}-2\gamma_{2}^{2}k_{u}k_{2}\right)}\label{eq:Rt_FT}
\end{equation}

\noindent with

\begin{eqnarray}
\epsilon^{2} & = & \delta_{\Vert}^{2}\cos^{2}\left(\alpha\right)+\delta_{\bot}^{2}\sin^{2}\left(\alpha\right)\nonumber \\
\gamma_{1}^{2} & = & \left(\delta_{\Vert}^{2}-\delta_{\bot}^{2}\right)\sin\left(\alpha\right)\cos\left(\alpha\right)\sin\left(\beta\right)\nonumber \\
\gamma_{2}^{2} & = & \left(\delta_{\Vert}^{2}-\delta_{\bot}^{2}\right)\sin\left(\alpha\right)\cos\left(\alpha\right)\cos\left(\beta\right)\nonumber \\
\mu_{1}^{2} & = & \delta_{\Vert}^{2}\sin^{2}\left(\alpha\right)\sin^{2}\left(\beta\right)+\delta_{\bot}^{2}\left[\cos^{2}\left(\beta\right)+\cos^{2}\left(\alpha\right)\sin^{2}\left(\beta\right)\right]\label{eq:coef2}\\
\mu_{2}^{2} & = & \delta_{\Vert}^{2}\sin^{2}\left(\alpha\right)\cos^{2}\left(\beta\right)+\delta_{\bot}^{2}\left[\sin^{2}\left(\beta\right)+\cos^{2}\left(\alpha\right)\cos^{2}\left(\beta\right)\right]\nonumber \\
\mu_{12}^{2} & = & \left(\delta_{\Vert}^{2}-\delta_{\bot}^{2}\right)\sin^{2}\left(\alpha\right)\sin\left(\beta\right)\cos\left(\beta\right)\nonumber 
\end{eqnarray}

\noindent and $\mathbf{k}_{v}\equiv k_{1}\mathbf{e}_{1}+k_{2}\mathbf{e}_{2}$.

We also note that because the depth of integration along the line
of sight is expected to be much larger then the turbulent correlation
lengths (i.e., $\Delta\gg\delta_{\Vert}$ and $\Delta\gg\delta_{\bot}$)
we have 

\begin{eqnarray}
\int e^{-\frac{1}{2}\left(\epsilon^{2}k_{u}^{2}+2\gamma_{1}^{2}k_{u}k_{1}-2\gamma_{2}^{2}k_{u}k_{2}\right)}\mathrm{sinc}^{2}\left(\frac{k_{u}\Delta}{2}\right)dk_{u} & \simeq & \int\mathrm{sinc}^{2}\left(\frac{k_{u}\Delta}{2}\right)dk_{u}\nonumber \\
 & \simeq & \frac{2\pi}{\Delta}.\label{eq:Rt(k)_int}
\end{eqnarray}

\noindent Inserting Equation (\ref{eq:Rt(k)_int}) in Equation (\ref{eq:FT})
with

\begin{equation}
H\left(\mathbf{k}_{v}\right)=e^{-\frac{1}{2}W^{2}k_{v}^{2}},\label{eq:beam_FT}
\end{equation}

\noindent and further using Equation (\ref{eq:Gauss_int}), it is
found that

\begin{equation}
\overline{B}_{\mathrm{t}}^{2}\left(\mathbf{k}_{v}\right)=\left\langle F_{0}^{2}\right\rangle \left\langle B_{\mathrm{t}}^{2}\right\rangle \left(2\pi\right)^{3/2}\frac{\delta_{\Vert}\delta_{\bot}^{2}}{\Delta}e^{-\frac{1}{2}\left[k_{1}^{2}\left(\mu_{1}^{2}+2W^{2}\right)+k_{2}^{2}\left(\mu_{2}^{2}+2W^{2}\right)-2\mu_{12}^{2}k_{1}k_{2}\right]}.\label{eq:Bt2(k)}
\end{equation}

\noindent Calculating the inverse Fourier transform of Equation (\ref{eq:Bt2(k)}),
still using Equation (\ref{eq:Gauss_int}), then yields

\begin{equation}
\left\langle \overline{\mathbf{B}}_{\mathrm{t}}\mathbf{\cdot}\overline{\mathbf{B}}_{\mathrm{t}}\mathbf{\left(\boldsymbol{\ell}\right)}\right\rangle =\frac{\left\langle F_{0}^{2}\right\rangle \left\langle B_{\mathrm{t}}^{2}\right\rangle }{N^{\prime}}e^{-\frac{1}{2}g\left(\boldsymbol{\ell};\delta_{\Vert},\delta_{\bot};\alpha,\beta\right)},\label{eq:Bt2(ell)}
\end{equation}

\noindent with

\begin{eqnarray}
g\left(\boldsymbol{\ell};\delta_{\Vert},\delta_{\bot};\alpha,\beta\right) & = & \left[\ell_{1}+\mu_{12}^{2}\ell_{2}/\left(\mu_{2}^{2}+2W^{2}\right)\right]^{2}/\left[\mu_{1}^{2}+\mu_{12}^{4}/\left(\mu_{2}^{2}+2W^{2}\right)+2W^{2}\right]\nonumber \\
 &  & +\ell_{2}^{2}/\left(\mu_{2}^{2}+2W^{2}\right)\label{eq:g}
\end{eqnarray}

\noindent and

\begin{equation}
\frac{1}{N^{\prime}}=\frac{\sqrt{2\pi}\delta_{\Vert}\delta_{\bot}^{2}}{\sqrt{\left[\mu_{1}^{2}+\mu_{12}^{4}/\left(\mu_{2}^{2}+2W^{2}\right)+2W^{2}\right]\left(\mu_{2}^{2}+2W^{2}\right)}\,\Delta}.\label{eq:1/N'}
\end{equation}

Following the treatment of \citet{Houde2009} for the ordered component
of the autocorrelation function we write

\begin{equation}
\left\langle \overline{B}_{0}^{2}\right\rangle \equiv\left\langle F_{0}^{2}\right\rangle \left\langle B_{0}^{2}\right\rangle \frac{\Delta^{\prime}}{\Delta},\label{eq:D'/D}
\end{equation}

\noindent where $\Delta^{\prime}\leq\Delta$ is the effective depth
over which the signal is integrated along the line of sight (in our
case approximately the thickness of the disk of M51), which is closely
related to the correlation length of the ordered polarized flux (see
Secs. 2.3 and 3.2 in \citealt{Houde2009}).

Combining Equations (\ref{eq:Bt2(ell)})-(\ref{eq:D'/D}) we find

\begin{eqnarray}
b^{2}\left(\boldsymbol{\ell}\right) & = & \frac{\left\langle \overline{\mathbf{B}}_{\mathrm{t}}\mathbf{\cdot}\overline{\mathbf{B}}_{\mathrm{t}}\left(\boldsymbol{\ell}\right)\right\rangle }{\left\langle \overline{\mathbf{B}}\mathbf{\cdot}\overline{\mathbf{B}}\left(0\right)\right\rangle }\nonumber \\
 & = & \left[\frac{\left\langle B_{\mathrm{t}}^{2}\right\rangle }{N\left\langle B_{0}^{2}\right\rangle +\left\langle B_{\mathrm{t}}^{2}\right\rangle }\right]e^{-\frac{1}{2}g\left(\boldsymbol{\ell};\delta_{\Vert},\delta_{\bot};\alpha,\beta\right)}\label{eq:b2(0)}
\end{eqnarray}

\noindent with the number of turbulent cells $N$ probed by the telescope
beam given by

\begin{equation}
\frac{1}{N}=\frac{\sqrt{2\pi}\delta_{\Vert}\delta_{\bot}^{2}}{\sqrt{\left[\mu_{1}^{2}+\mu_{12}^{4}/\left(\mu_{2}^{2}+2W^{2}\right)+2W^{2}\right]\left(\mu_{2}^{2}+2W^{2}\right)}\,\Delta^{\prime}}.\label{eq:1/N}
\end{equation}

If we choose to align one of the axes of the observer's coordinate
system on the plane of the sky (i.e., the $v_{2}$-axis) with the
large scale magnetic field $\mathbf{B}_{0}$, then $\beta=0$ and
Equations (\ref{eq:b2(0)}) and (\ref{eq:1/N}) reduce to Equations
(\ref{eq:b^2_inclined}) and (\ref{eq:N_inclined}), respectively.

\end{document}